\documentclass[conference]{IEEEtran}

\newif\ifsubmission
\submissionfalse % submission version

\usepackage{graphicx}
\usepackage{subfigure}
\usepackage{multirow}
\usepackage{multicol}
\usepackage{threeparttable}
\usepackage{amsmath}
\usepackage{color}
\usepackage{colortbl}
\usepackage{bm}
\usepackage{amssymb}
\usepackage{times}

\definecolor{blue}{rgb}{0,0,1}
\definecolor{Gray}{gray}{0.9}
\definecolor{Orange}{rgb}{1,0.5,0}

\newcommand{\sys}{LookAhead}

\begin{document}

%\title{LookAhead: Augmenting Crowdsourced Website Reputation Systems by Prediction of Eventual Rating}
\title{LookAhead: Augmenting Crowdsourced Website Reputation Systems With Predictive Modeling}
\vspace{-2cm}

\author{\IEEEauthorblockN{Sourav Bhattacharya$^1$, Otto Huhta$^2$, and N. Asokan$^{2,3}$}
\IEEEauthorblockA{
$^1$Bell Laboratories, Ireland, $^2$Aalto University, Finland, $^3$University of Helsinki, Finland\\
Email: sourav.bhattacharya@bell-labs.com, otto.huhta@aalto.fi, asokan@acm.org}}

% make the title area
\maketitle

%%%%%%%%% Abstract %%%%%%%%
%!TEX root = ../submission.tex
\begin{abstract}

Unsafe websites consist of malicious as well as inappropriate sites, such as those hosting questionable or offensive content. Website reputation systems are intended to help ordinary users steer away from these unsafe sites. However, the process of assigning safety ratings for websites typically involves humans. Consequently it is time consuming, costly and not scalable.  This has resulted in two major problems: (i) a significant proportion of the web space remains unrated and (ii) there is an unacceptable time lag before new websites are rated. 

In this paper, we show that by leveraging structural and content-based properties of websites, it is possible to reliably and efficiently predict their safety ratings, thereby mitigating both problems. We demonstrate the effectiveness of our approach using four datasets of up to 90,000 websites. We use ratings from Web of Trust (WOT), a popular crowdsourced web reputation system, as ground truth. We propose a novel ensemble classification technique that makes opportunistic use of available structural and content properties of webpages to predict their eventual ratings in two dimensions used by WOT: trustworthiness and child safety. Ours is the first classification system to predict such subjective ratings and the same approach works equally well in identifying malicious websites.  Across all datasets, our classification performs well with average F$_1$-score in the 74--90\% range. 
%Depending on the specific use case, our system can be tuned to minimize false positives or false negatives.

%,  outperforming a recently proposed approach that predicts safety ratings using only structural properties.  

\end{abstract}

%%%%%%%% Introduction %%%%%%%
%!TEX root = submission.tex
\section{Introduction}
\label{sec:intro}

\ifsubmission
Internet scammers
\else
The Internet has revolutionized the way we communicate today and has
already become an integral part of our daily lives. The immense
popularity of the Internet, with an increasing user base of billions,
has naturally attracted miscreants.
They 
\fi
set up various types of ``unsafe'' websites to lure their
victims. These include \emph{malicious} sites,
intended for {\em phishing}, {\em drive-by-downloads} of malware etc. as well as sites that are \emph{inappropriate} in some sense.
Examples include websites hosting offensive, objectionable, hateful or illegal content,
and misusing private user data.

%% Changed 25.11.:
%Today, the Internet consists of over a billion websites
%with hundreds of thousands of new sites being set up each day.
%As anyone is able to set up a website, it can be difficult for
%a user to know which ones are safe to visit. Some websites are
%outright \textit{malicious} intended for phishing or drive-by-downloads
%whereas others can be just \textit{inappropriate} for a certain
%group of users. Examples of this include hosting offensive or illegal
%content, or misusing private user data.
%% --------------

A variety of mechanisms have been developed for steering unsuspecting
users away from unsafe websites. Popular browsers present interstitial
security warnings when users attempt to navigate to a known malicious
website~\cite{AF13}. Several anti-virus vendors maintain website
reputation systems (e.g.,
TrustedSource\footnote{http://www.trustedsource.org/}. These
systems use a combination of machine learning techniques and manual
expert evaluation to arrive at the rating for a given website. A
popular sub-category of reputation systems are those that make use of
input ratings that are \emph{crowdsourced} from the users of the
system. \emph{PhishTank}\footnote{http://www.phishtank.com/} and
\emph{Web of Trust} (WOT)\footnote{https://www.mywot.com/} are
examples of web reputation systems that rely fully or partly on
crowdsourced ratings. An advantage of crowdsourced ratings is that
the ratings can cover a broader class of unsafe websites,
including those that are perceived to be inappropriate but not
outright malicious~\cite{CK11}. 
\ifsubmission
\else
Typically these rating systems are
queried by dedicated browser extensions which can signal the result to
the user in the form of a color-coded glyph, e.g., a red glyph
indicating an unsafe site and a green glyph indicating a safe site (see Section~\ref{sec:background}).
\fi

\begin{figure}[!t]
\centering
\includegraphics[width=0.95\linewidth]{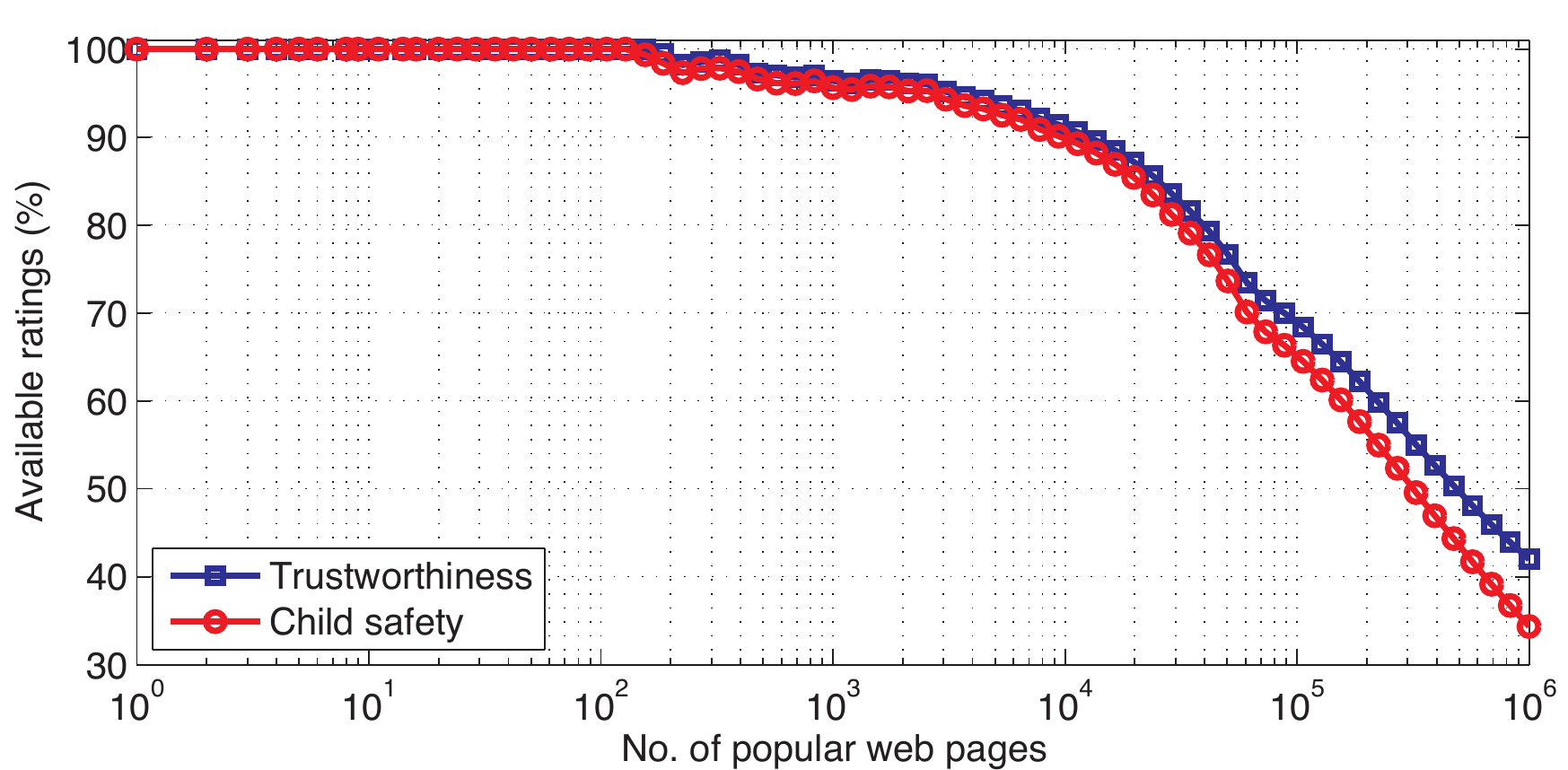}
\vspace{-3mm}
\ifsubmission
\caption{Cumulative availability (\%) of WOT ratings (trustworthiness and child safety) for the one million most popular webpages, July 2014.}
\else
\caption{Cumulative availability (\%) of WOT reputation ratings along two dimensions (trustworthiness and child safety) for the one million most popular webpages as of July 2014 collected from Alexa (http://www.alexa.com). The plot highlights that often (over $58\%$) webpages lack reputation ratings in at least one dimension to indicate their safety.}
\fi
\label{fig:motivation}
\end{figure}

All website reputation rating systems, especially those that
involve humans in the rating process, suffer from two major disadvantages:
\begin{itemize}
\itemsep0em
\ifsubmission
%% Submission version
\item \emph{Insufficient coverage}: Rating coverage is limited. Fig.~\ref{fig:motivation}, shows the availability of WOT reputation ratings ({\em trustworthiness} and {\em child-safety}) for the one million most popular webpages. The majority of the pages are unrated.
\item \emph{Time lag}: The time gap between a new website coming
  online and the system assigning a rating can be in the order of days to months.
\else
%% Full version
\item \emph{Insufficient coverage}: Often the webspace coverage by a reputation system is limited. This is due to the high cost and poor scalability of obtaining experts' ratings, as well as the lack of motivation for users to participate actively in rating webpages. This problem is illustrated in Fig.~\ref{fig:motivation}, which shows the availability of WOT reputation ratings for the one million most popular webpages.
In WOT websites are rated according to two dimensions, {\em trustworthiness} and {\em child safety}, which are both subjective. A majority of the webpages are unrated: $58\%$ for trustworthiness and $66\%$ for child safety.

%As a corollary, a large proportion of websites tend to remain unrated. For example,Figure~\ref{fig:motivation} shows the availability of WOT reputation ratings for the one million most popular web pages. A majority of these web pages are unrated.
%\itemsep0em
\item \emph{Time lag}: The time gap between a new website coming
  online and the system assigning a rating to it can be long. Often
  the gap can be in the order of days to months. This is particularly
  problematic because unsafe websites tend to be short-lived, with
  lifetimes often in the order of hours or days~\cite{MC09}. 
%{\color{blue} During this period, a large number of users are exposed
%to unsafe websites.}  
\fi
\end{itemize}
A consequence of these drawbacks is that users who rely on such reputation systems to protect them from unsafe contents remain vulnerable when many unsafe websites are unrated. 
\ifsubmission
\else
Regardless of the concerns as to whether such reputation systems based on crowdsourced ratings are effective~\cite{tyler08evaluating}, the sheer number of users who rely on such systems (see Section III) warrants solutions to mitigate this vulnerability.

Rating systems that make use of machine learning techniques, provided that they are fast, can address these shortcomings. 
\fi
Machine learning has already been extensively used for detecting malicious websites based on the structure and content of web pages \cite{canali11prophiler, cova2010detection, rieck2010cujo}.
It is plausible that similar techniques can also be applied to predicting
the future rating of a website in subjective dimensions as used in systems like WOT.
Thus, our work addresses
the following research question:
\emph{Can we reliably predict the eventual rating of an unrated website?}

In this paper, we describe LookAhead, the system we built in the process of investigating this question. LookAhead uses a combination of structural and content-based features to predict the eventual rating a website is likely to receive. In reality, not all feature types can be extracted from all webpages (see Section~\ref{sec:predictiveModeling}) and to mitigate this feature unavailability problem we propose an ensemble classification approach. Accordingly, we train different classifiers for each feature type and present different combination strategies to estimate the overall rating. For the structure of the websites, we consider HTML and JavaScript-based features. However, we show that structural features alone would not be sufficient for accurate predictions. Therefore, we introduce a novel content-based feature set, that is extracted from the malicious outward links and the text present on a webpage.

%also consider the content of the website in the form of malicious outward links and the topics present. 

%A previously proposed
%system, Prophiler~\cite{canali11prophiler}, can be used to accurately estimate
%the safety of a web page based only on its structural properties, such as
%the HTML and Javascript contents.
%Crowdsourced reputation systems alleviate the scalability problem of an expert-rated %system, but
%they introduce a significant variability in the form of user assigned ratings.
%This variability makes the learning task difficult and non-trivial.
%Contrary to existing research, which mainly relies on expert-rating of
%maliciousness, in this paper we address the following research question: \emph{Can we %reliably predict the eventual crowdsourced rating of an unrated website?}
%We employ WOT as our target web reputation system, which is shown to be more accurate in %its ratings compared to other similar systems~\cite{CK11}. WOT provides ratings in two %dimensions, namely trustworthiness and child safety, and covers both maliciousness and %inappropriateness aspects of web page domains.

We make the following contributions:
\begin{itemize}
%\item A \textbf {large-scale study} of the effectiveness of using only
%  structural features (as used in Prophiler~\cite{canali11prophiler}). We show that they are not as effective in identifying unsafe
%  websites in general (as defined by WOT ratings) as they were in
%  identifying malicious websites. We use a much larger dataset than that was
%  used in~\cite{canali11prophiler} (Section~\ref{ssec:prophiler}).
\item \textbf{Content-based features} for effectively predicting the future rating of a website.
In particular, we propose a novel use of the {\em
    empirical cumulative distribution function} (ECDF) as a feature
  set to extract clues about the content of a web page based
  on ratings of hyperlinks embedded in it
  (Section~\ref{sssec:ecdf}). We also propose how \emph{topic
    modeling} techniques can be used to extract features that capture
  the theme of a webpage (Section~\ref{sssec:topic}).
\item \textbf{\sys{}: an adaptive ensemble classification technique} 
  effectively combining several classifiers for structural and content-based feature sets by learning combination
  weights from data (Section~\ref{ssec:classifier}). 
%Depending on the situation the system operator can vary the combination strategy to achieve either a low false positive rate or a low false negative rate.
%\item \textbf{LookAhead}, our own proposed rating system that combines structural and content-based features with the adaptive hierarchical classification technique.(Section~\ref{sec:predictiveModeling})
\item \textbf{Systematic evaluation} of LookAhead on
several datasets with up to $90,000$ web pages. (Section~\ref{sec:experiments})
We also evaluate the performance of Prophiler {\cite{canali11prophiler}, which uses only structural features, on the those datasets.
We show that the
  performance is significantly (statistically) better than using only
  the structural features of web pages, as in Prophiler} (Section~\ref{sec:eval}).
In particular, this holds across both subjective dimensions (trustworthiness and child safety), as well as maliciousness.

\end{itemize}

\section{Related Work}
\label{sec:related}

A typical approach for helping users avoid malicious websites is to make use of blacklists of known bad websites. For example, Microsoft's Internet Explorer and Mozilla Firefox web browsers warn users when they try to visit a page present on a blacklist. Unfortunately, blacklists suffer from a number of shortcomings, e.g., blacklists are required to be updated periodically, are often slow to reflect new malicious websites, and have poor coverage of malicious webspace. Moreover, adversaries often try to masquerade malicious webpages as benign by making slight modifications to their URLs. To mitigate problems with blacklists, Felegyhazi {\em et al.} propose a system that, given an initial blacklist of domains, tries to predict potentially malicious domains based on nameserver features and registration information~\cite{Felegyhazi10}. Prakash {\em et al.} propose five different heuristics that allow synthesizing new URLs from existing ones. The authors use this idea to enlarge the existing blacklist of malicious URLs~\cite{Prakash10}.

Going beyond blacklists, application of machine learning techniques to successfully identify malicious websites has become popular. Ma {\em et al.}~\cite{Ma09BeyondBlacklists} explore the use of lexical features, including the length and number of dots in URLs, host-based features, such as IP address, domain name and other data returned by WHOIS queries~\cite{daigle2004whois} to identify malicious web links. They evaluated their approach using  20,000 URLs drawn from different sources, specifically benign URLs are collected from DMOZ Open Directory Project\footnote{http://www.dmoz.org/ [Retrieved:\today]}  and Yahoo's directory\footnote{http://random.yahoo.com/bin/ryl/ [Retrieved: \today]}, and malicious URLs from PhishTank and Spamscatter~\cite{anderson2007spamscatter}. They reported a {\em false positive rate} (FPR) of $14.8\%$ and a {\em false negative rate} (FNR) of $8.9\%$.

Another popular approach is to analyze the structural properties of webpages, especially looking for known malicious patterns within the embedded javascript, to identify malicious sites that trigger drive-by-download attacks~\cite{cova2010detection,curtsinger2011zozzle,FP07,LJO09,rieck2010cujo}.
JSAND by Cova {\em et al.} \cite{cova2010detection} combines anomaly detection with emulation and uses a naive Bayes
classifier. Out of around 800 malicious pages and scripts, they report a
very low FNR of 0.2\%, although they do not report the corresponding FPR.
Cujo \cite{rieck2010cujo} by Rieck et al. considers both
static and dynamic javaScript features and classifies websites using
Support Vector Machines (SVM). They look at around 220,000 benign websites
and 600 drive-by-download attacks and report an average {\em true positives rate} (TPR) of 94.4\% with a 0.002\% FPR.
They also ran JSAND on the same dataset and
report a 99.8\% TPR. Finally, ZOZZLE by Curtsinger et al.
\cite{curtsinger2011zozzle}
considers over 1.2 million javascript samples and achieves FPR and
TPR in the range of 1.2-5.1\%.

Closest to our work is Prophiler \cite{canali11prophiler} by
Canali et al., which also looks at identifying malicious websites,
but using only static features related to the URL and the structure of a page.
For each web page they consider $37$ URL-based, $20$ HTML and $26$ JavaScript
features and train three different classifiers, one for
each feature type. They reported 0.77\% false negatives and 9.88\% false positives\footnote{Not to be confused with FNR or FPR; see Appendix~\ref{app:fpr} for exact definitions.} with a dataset of 15,000 web pages.

%Although there have been several previous attempts at using machine learning techniques to detect malicious webpages based on their structure, Prophiler by Canali et al. appears to be the one with the best performance~\cite{canali11prophiler}. In addition, all the features used are ``static features'' in that they do not require rendering the page or running any embedded code. The main design objective of Prophiler is to quickly classify a web page as either potentially benign or malicious. If a webpage is identified as potentially malicious by the front-end filter, it can then be further analyzed with a resource-intensive dynamic analysis system, e.g., Wepawet~\cite{cova10detection}.

Below we summarize the main differences between our work and previous research.
%We can identify a few main differences between our work, and that done previously in this area.
While systems like Prophiler \cite{canali11prophiler}
and JSAND \cite{cova2010detection} report good results for detecting
malicious websites, we consider a much broader and non-trivial problem of
predicting subjective rating dimensions like trustworthiness and child-safety of a website.
In addition, we consider not only
the structure and URL of a web page but also the content presented on
that page.
To our knowledge,
this is the first study combining static and content-based website
features to predict the future reputation rating of a website.

\section{Web Reputation System WOT}
\label{sec:background}

\begin{figure}[!t]
\centerline{
	\subfigure[]{
	\label{fig:framework:1st}
	\includegraphics[width=0.6\linewidth]{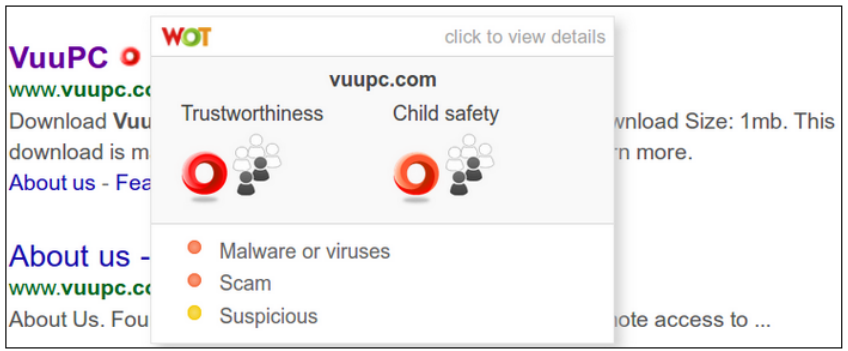}\label{fig:wot}}
	\subfigure[]{
	\label{fig:framework:2nd}	
	\includegraphics[width=0.27\linewidth]{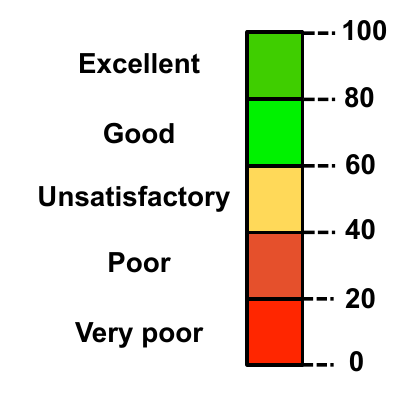}\label{fig:wot_score}}	
	}
\vspace{-3mm}	
\caption{ (a) WOT user interface showing aggregated user ratings of a
  web page being poor (red) in both dimensions (trustworthiness and
  child safety). The confidence in
  the rating is represented by the number of dark figurines. (b) WOT
  divides the reputation rating range into five levels; The ratings of a
  webpage are indicated by a color-coded glyph representing the level.}
\label{fig:rsDistribution}
\end{figure}

WOT is a web reputation system, that provides reputation ratings of the domain of a given URL in two dimensions, trustworthiness and child safety, as an integer in the range [0-100]. WOT builds the reputation ratings of a web domain mainly based on crowdsourced input ratings from a large user base and then applying a proprietary aggregation algorithm. Additionally, WOT uses input from other {\em trusted sources}, but the identities of these sources are not public.

The user front-end of WOT is a browser extension that scans the page being rendered in the browser for URLs, looks up their reputation
ratings in the WOT backend, and shows the results as color-coded glyphs. For example, Fig.~\ref{fig:wot} shows a red glyph next to a website
deemed unsafe by WOT. The rating space is divided into five levels,
with a color code assigned to each level, see Fig.~\ref{fig:wot_score}. Clicking on the glyph brings out a pop-up window that shows
more information about the rating. WOT's confidence in a rating
(which we believe correlates with the number of users who had given
input ratings) is also indicated in this window, represented by a set
of dark figurines (up to five). WOT has seen well over 100 million\footnote{http://mywot.net [Retrieved: \today].}
downloads. It is also used by large scale services like {\em Facebook} and
{\em Mail.ru}. It is reasonable to assume that the user base of WOT and
similar rating systems runs into tens of millions.

Our objective is to see if we can use information found on a hitherto unrated webpage to predict what rating it will receive. In this paper we use WOT as the target reputation system. However, our proposed method is generic and would work with any web reputation system. 
We therefore use existing WOT ratings as the ground truth, and apply a supervised learning-based algorithm for model learning.
Moreover, instead of building a regression model, we formulate the web page reputation prediction as a binary classification task~\cite{bishop07pattern}. The binary classification approach divides the reputation ratings into two (coarse) groups by applying a suitable threshold on the reputation ratings. This approach helps to minimize the effect of subjective variations among users in their ratings. Given a reputation rating $r \in [0,100]$ of a URL, the class information of the URL is computed using the following simple rule:

\begin{equation}
class(r) = 
\begin{cases}
bad & \mbox{if}\,\,\,  r < \mathcal{T}_h\\
good & \mbox{otherwise} 
\end{cases}
\label{eqn:classification}
\end{equation}

\noindent In our experiments (Section~\ref{sec:eval}) we present results for $\mathcal{T}_h = 40$, while Appendix~\ref{app:fpr} contains results for $\mathcal{T}_h = 60$.

%%% Methodology / Approach
%!TEX root = submission.tex
\section{\sys: Predicting Safety Ratings}
\label{sec:predictiveModeling}

\begin{figure}[!t]
\centering
\includegraphics[width=0.65\linewidth]{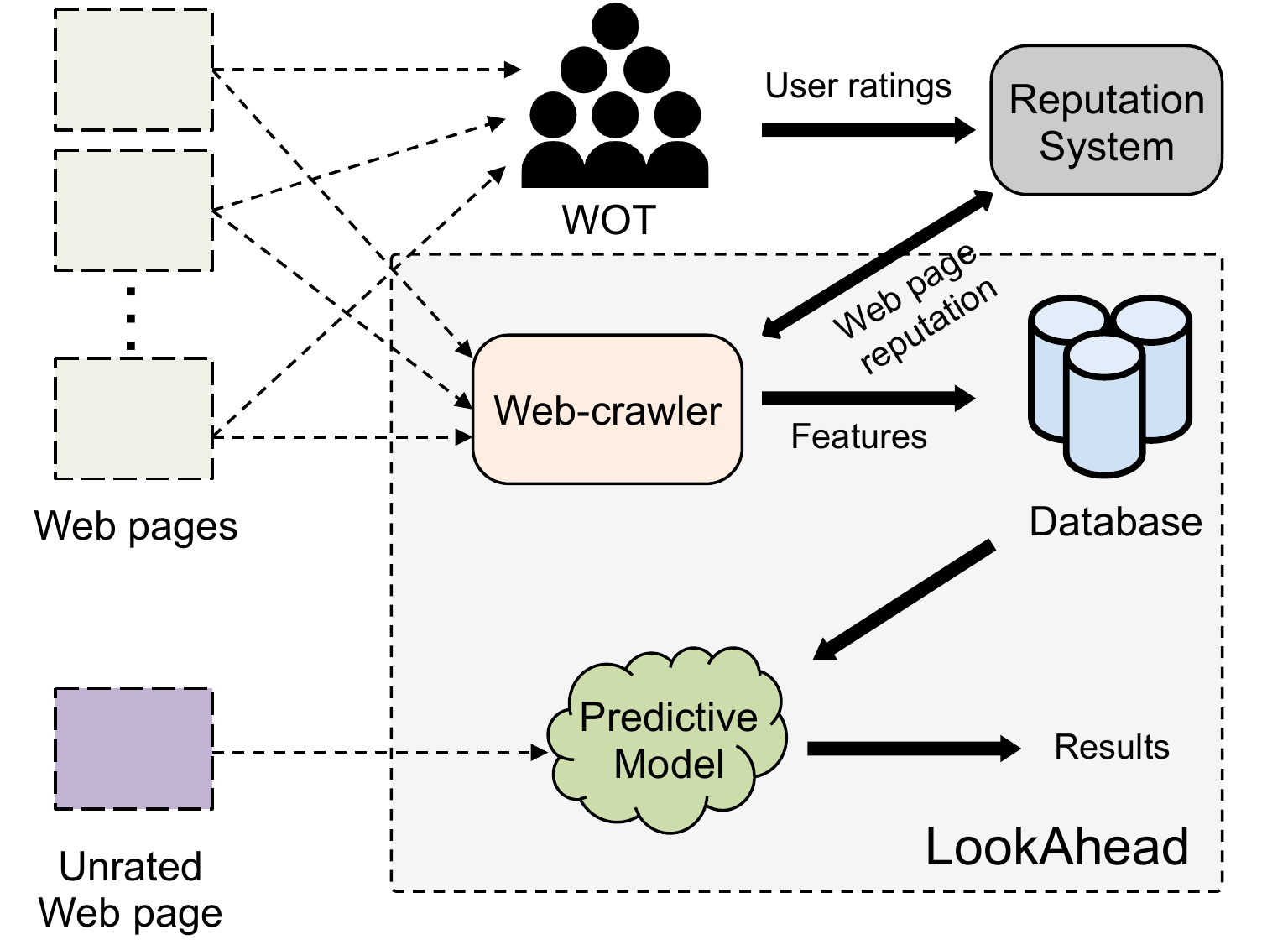}
\vspace{-3mm}
\caption{An architectural overview of \sys{} in association with a crowdsourced web reputation system WOT.}
\label{fig:system}
\end{figure}

The main objective of \sys{} is to extend the existing capabilities of a web reputation system, such as WOT, by predicting the eventual ratings of previously unrated websites. The predictive approach utilizes existing reputation ratings of a large number of webpages to learn a mapping function from various webpage features to a set of target classes, in our case, either {\em good} or {\em bad} (see Equation~\ref{eqn:classification}). Fig.~\ref{fig:system} illustrates an overview of our web safety prediction approach combined with WOT. The \sys{} part, highlighted in the figure, is composed of a web crawler, a database, and a predictive model. The web crawler extracts various features from webpages and stores them, along with reputation ratings in various dimensions\footnote{Reputation ratings in trustworthiness and child safety are obtained using  API calls to the WOT system. For details see https://www.mywot.com/wiki/API [Retrieved: \today].}, to a database. The predictive model is responsible for learning a classification model and also responsible for predicting web safety of unrated URLs.

%{\color{blue}Depending on the application, we may wish to tune our classifier towards low false negatives or low false positives. The former is usually desirable for security, i.e. in order to detect as many bad websites as possible, but on the other hand it results in poorer usability as more good websites are being incorrectly classified as bad. We do not want to pin down the exact use case for LookAhead and so we allow the user or system administrator to tune the classifier for either a low FNR or FPR by adjusting a decision threshold for class prediction probabilities returned by the Random Forest classifier we use. This threshold is a value between 0 and 1 and represents the level of certainty the classifier has for a given prediction.}

% ALTERNATIVE:
%The goal for LookAhead is to predict the reputation of a web page when
%no prior ratings are available, and so we target a low False Positive rate, and
%limit the number of pages that are incorrectly classified as bad. This way, we
%accept the fact that we will not identify all bad websites, but argue that it is
%better than the current situation where the users have no information for these
%websites. Conversely, having a high False Positive rate would mean users would
%be less likely to pay attention to any warnings based on LookAhead due to previous
%incorrect predictions.}

\ifsubmission

\else
The efficacy of \sys{} relies mainly on the quality of the learned predictive model, as well as its generalizability to unforeseen URLs. As in any supervised machine learning approach, precursor to the model learning is a time consuming bootstrapping step~\cite{bhattacharya14using}. A typical bootstrapping process involves obtaining suitable feature representation of the data, in our case websites, and collecting accurate ground truth labels. Often the feature extraction process is referred to as feature-engineering, as it relies on domain-specific expertise. The optimal feature set is often dependent on the target classes of interests and obtaining an optimal features set has been identified as a non-trivial problem~\cite{liu05toward}.
\fi
In this work we consider two types of features to represent websites: (i) {\em structural} features, which are extracted from the HTML and embedded JavaScript codes, and (ii) {\em content} features that capture ratings of web links embedded in the page and the thematic structure of page text.

% Changed 25.11.:
%We extract features from web pages and use a classfication approach, a well established supervised learning-based algorithm~\cite{bishop07pattern}, to train a recognizer for automatically determining the safety rating of web pages. Our methodology relies on extracting suitable features that capture properties of a web page. For {\em ground-truth} of web safety, we rely on crowdsourced based reputation ratings of web pages, which we obtain through API calls to the WOT reputation system. Figure~\ref{fig:system} illustrates an overview of our classification framework in general. Accordingly, we have developed a web crawler that extracts different kinds of features and reputation ratings (trustworthiness and child safety) for a large number of web pages and stores the data in a database to be used by machine learning algorithms for building a predictive model (recognizer). In the following we describe our recognition framework, LookAhead, in detail.

%A recognizer employing a supervised machine learning approach requires to perform a time consuming bootstrapping step before it can be deployed~\cite{bhattacharya14using}. A typical bootstrapping process involves obtaining suitable feature representation of the data and collecting a large amount of ground truth information. Often the feature extraction process is referred to as feature-engineering as it relies on domain-specific expertise. Moreover, the optimal feature-set is dependent on the target classes of interests and has been identified as a non-trivial problem~\cite{liu05toward}. 

\subsection{Structural Features of Web Pages}
\label{ssec:features}

For structural features, we mainly rely on past research that has identified and successfully validated a large set of features, based on HTML and embedded JavaScripts codes, to identify malicious webpages. Specifically, we adopt the handcrafted and domain specific features proposed by Canali {\em et al.}, as part of their Prophiler system~\cite{canali11prophiler}. In the evaluation section (see Section~\ref{ssec:prophiler}), we consider Prophiler as our main baseline algorithm.

%We use the HTML and javascript features from Prophiler~\cite{canali11prophiler} as the starting point. Unfortunately we were not able to take advantage of the URL features used by Prophiler since none of the records in our dataset had all of the URL features available, and a large majority (82\%) had less than half of them present. The two feature sets we used were:

\subsubsection{HTML-based Features}
\label{sssec:html}
We adopt the same $20$ HTML features\footnote{Readers are referred to~\cite{canali11prophiler} for an exhaustive and in-depth description of all the HTML features.} used by Prophiler. Examples include the number of {\em iframe} tags, the number of hidden elements, the number of script elements, the percentage of unknown tags, and the number of malicious patterns, e.g., presence of the $meta$ tag~\cite{canali11prophiler}.    

\subsubsection{JavaScript-based Features}
\label{sssec:js}
We use the same $24$ JavaScript-based features used by Prophiler, which are extracted by analyzing either the JavaScript file or the $<$script$>$ element embedded within the HTML text. Examples of JavaScript-based features include the number of times the {\em eval()} function is used, the number of occurrence  of the {\em setTimeout()} and {\em setInterval()} functions, the number of {\em DOM} modification functions, and the length of the script in characters (see~\cite{canali11prophiler}).

\subsection{Content Features of Web Pages}
\label{ssec:contentFeatures}

Contrary to the state-of-the-art approaches, in this paper we propose the use of a novel set of features based on (1) {\em empirical cumulative distribution function} (ECDF) of the reputation ratings of embedded forward links and (2) {\em topic modeling}. The main intuition behind using these features is that by learning (unsupervised) webpage content properties, we avoid the need for handcrafted features based on domain knowledge. In our evaluation, we show that the proposed novel features improve the recognition performance significantly (see Section~\ref{sec:eval}).             

%The main benefits of using ECDF and topic model-based features are: (i) no requirement of domain knowledge to construct suitable feature representation of a web page and (ii) the features capture properties related to the content and theme of web pages. 

% Added 25.11.:
%We expect that the structural features of a web page are not sufficient for predicting the safety of a website and thus we wish to also capture content related properties of the web page. For this we propose two new features: embedded (forward) links and a topic model-based feature-set. In the following, we will describe both methods in detail.
% -------------
%Towards capturing the content related properties, we propose the use of features extracted from the embedded (forward) links of web pages. To gain further insights into the thematic content, we also propose the use of a topic model-based feature-set. In the following we describe each type of features in detail.

\subsubsection{Embedded Link-based Features} 
\label{sssec:ecdf}

To extract simple yet effective clues about the content of a web page, we hypothesize that the content of a page is related to
the content of the pages it links to.  In other words, we make use of the hypothesis: ``{\em You are the company you keep}". This saying is based on the fact that often knowledge about a unknown person's friends provides some idea about the person's interests or personality. Similar ideas have been successfully applied in recommender systems~\cite{breese98empirical} and in detecting susceptibility of mobile devices for malware infections~\cite{truong14company}.

%the field of collaborative-filtering to recommend products to a person based on the interests of people similar to him . Recently, Truong {\em et al.} have successfully applied the same hypothesis in predicting malware infection susceptibility of a mobile device  by looking at the set of applications running on the device.

\begin{figure}[!t]
\centering
\includegraphics[width=0.9\linewidth]{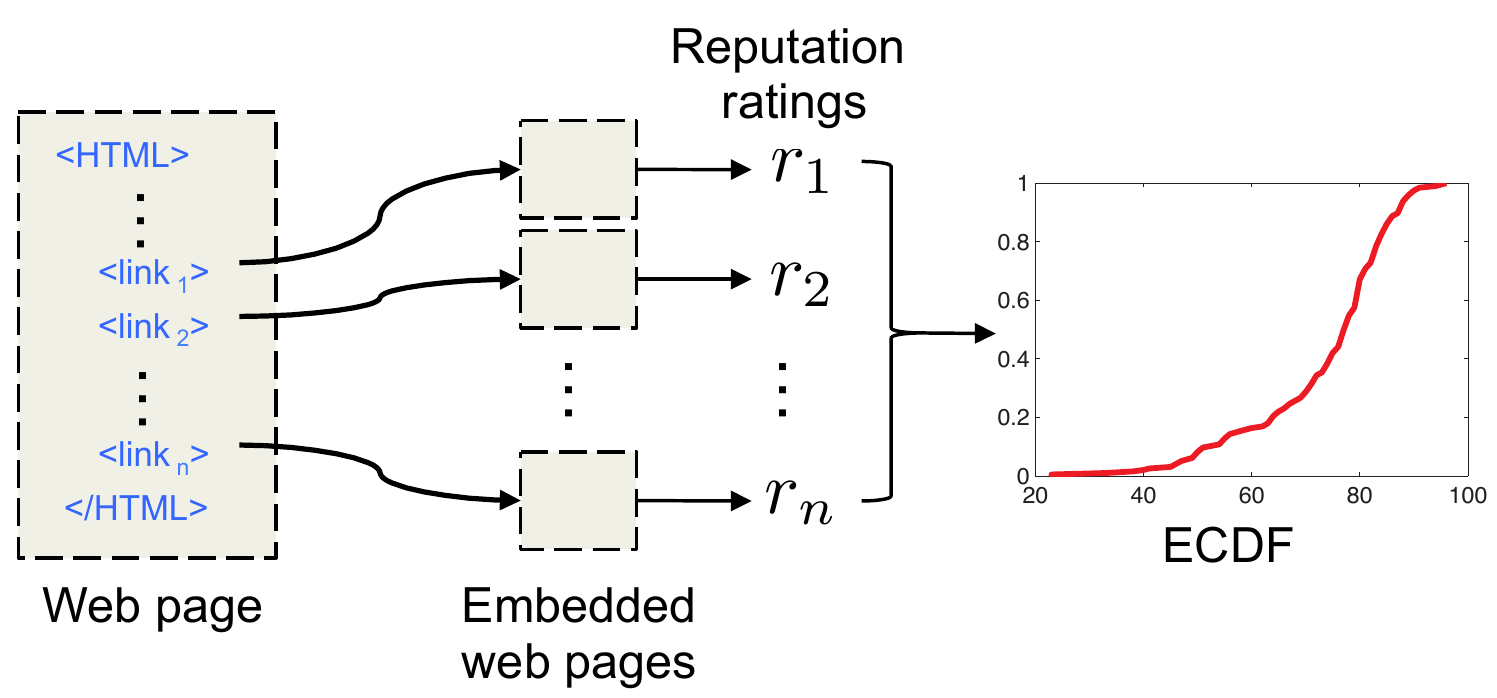}
\vspace{-3mm}
\caption{Overview of the embedded link-based feature extraction procedure. For a given webpage (on the left), we identify the set of out going web links (in the middle) present on the page and fetch the reputation ratings of those embedded links. We use the reputation ratings to compute an empirical cumulative distribution function (ECDF).}
\label{fig:ecdfOverview}
\end{figure}

%In the case of web pages, the hypothesis can be easily conceptualized. 
%For example, often a web page contains a set of links to other web pages. The reputation ratings of the linked web pages can be used to extract important clues about the content rating of the web page under consideration. 
Building on this idea, we propose a feature extraction scheme utilizing the available reputation ratings of embedded links. However, web pages may contain an arbitrary number of embedded links, e.g., the number can vary between zero and a very large number (few hundreds). Moreover, the range of the reputation ratings can be arbitrary. Thus we need a feature representation scheme that can compactly represent an arbitrary number of outgoing links, while remaining robust in the face of arbitrary ranges of ratings. 

%Thus, a novel feature extraction scheme is needed that can handle the arbitrary number of links, at the same time it is less sensitive to the arbitrary ranges of the reputation ratings. 

\begin{figure*}
\centerline{
	\subfigure[Histogram showing the frequency distribution of reputation ratings of embedded web links found within a web page.]{
	\label{fig:framework:1st}
	\includegraphics[width=0.3\linewidth]{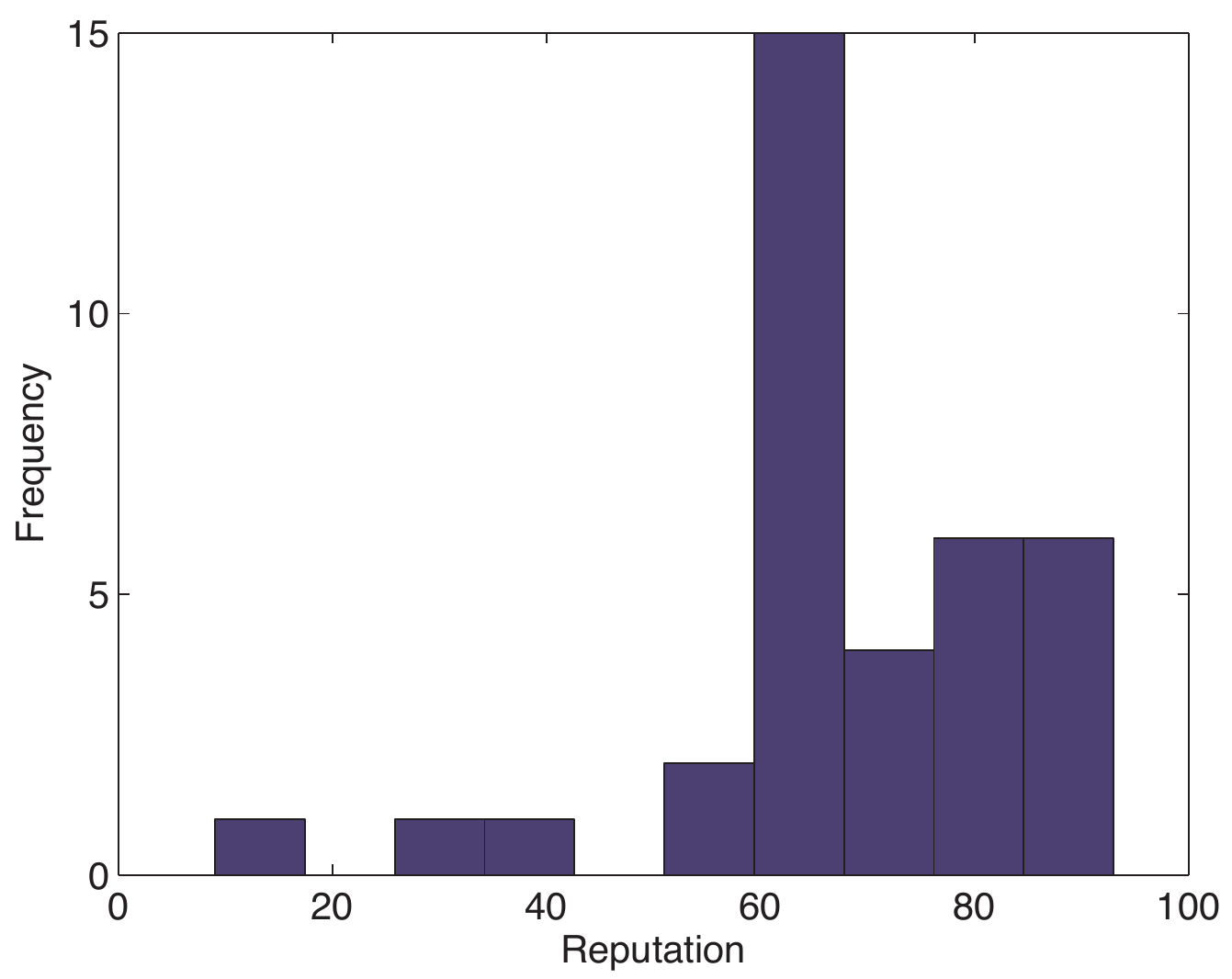}\label{fig:histRS}}
	\subfigure[Empirical Cumulating Distribution Function (ECDF) computed from the reputation ratings.]{
	\label{fig:framework:2nd}	
	\includegraphics[width=0.3\linewidth]{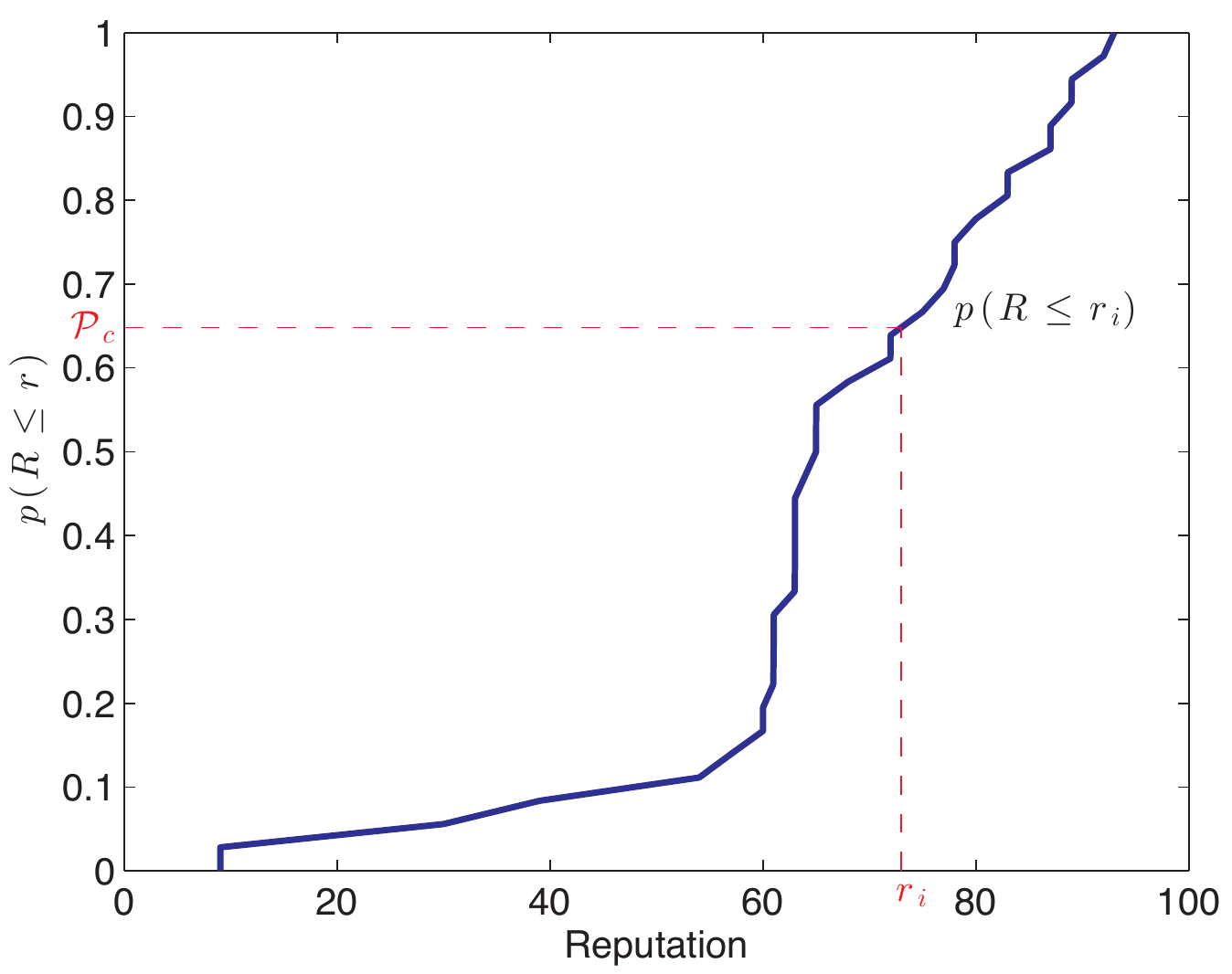}\label{fig:ecdfExample}}
	\subfigure[The ECDF-based features by computing the inverse.]{
	\label{fig:framework:1st}
	\includegraphics[width=0.292\linewidth]{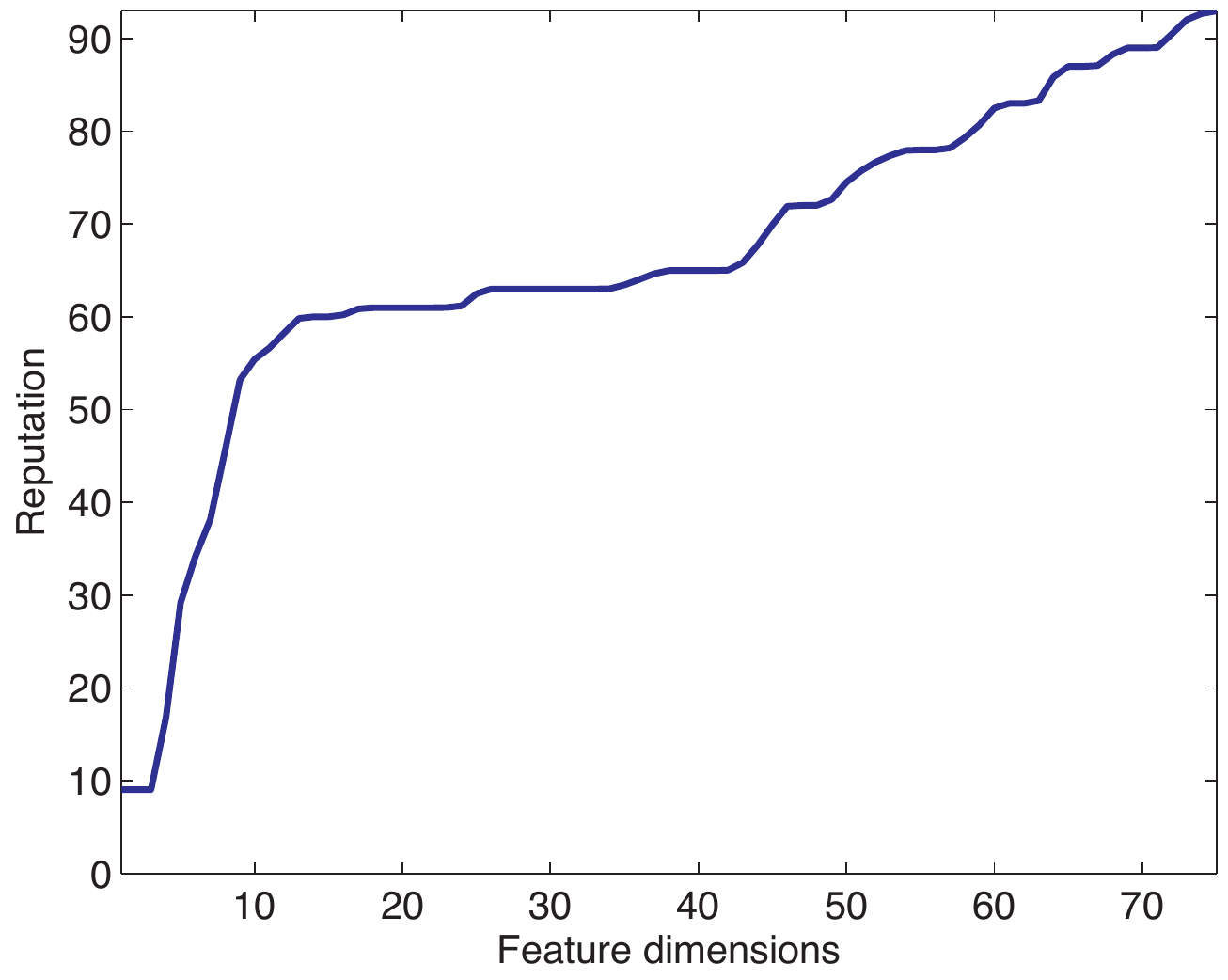}\label{fig:ecdfFeatures}}
	}
\vspace{-3mm}	
\caption{Exemplary illustration of (a) the distribution of reputation ratings, (b) their Empirical Cumulative Distribution Function, and (c) ECDF-based features.}
\label{fig:rsDistribution}
\end{figure*}

ECDF-based feature extraction has been previously explored in the field of ubiquitous computing and mobile sensing to represent human motion characteristics from continuous accelerometer data streams~\cite{bhattacharya14using,hammerla13preserving,plotz11feature}. However, the method has attracted very little attention outside the sensing domain. The simplicity and fast computation time of the ECDF features make it a viable option for using it in static web page analysis. Contrary to mobile sensing, in this paper we primarily focus on discrete reputation ratings. Fig.~\ref{fig:ecdfOverview} shows a pictorial overview of the basic idea of extracting ECDF-based features from a set of embedded web links. 

More formally, let $R = \{r_1, r_2, \ldots, r_n\}$ denotes the set of available reputation ratings of all the embedded web links on a page, where $r_i \in \mathbb{I}_{[0, 100]},\, \forall i \in \{1,\ldots,n\}$. The ECDF $\mathcal{P}_c(r)$ of $R$ can be computed as below:
\begin{eqnarray}
\mathcal{P}_c(r) = p(X \le r),\label{eqn:ecdf}
\end{eqnarray}
where, $p(X = r)$ is the probability of observing an embedded web link with a reputation rating of $r$ and $X$ is a random variable that takes values from $R$ (uniformly at random). For example, Figure~\ref{fig:histRS} shows an exemplary histogram of reputation ratings of web links found within a web page and Figure~\ref{fig:ecdfExample} shows the corresponding ECDF computed using Equation~\ref{eqn:ecdf}. Note that $\mathcal{P}_c(r)$ is defined on the entire range of the reputation ratings for embedded web links and is a monotonically increasing function. 

Often the distribution of reputation ratings for embedded links is multimodal, e.g., as in our example shown in Figure~\ref{fig:histRS}. In order to learn from such distributions, a recognition system should extract descriptors that relate to the shape and spatial position of the modes~\cite{hammerla13preserving}. The shape of the distribution is captured as $\mathcal{P}_c$ increases from $0$ to $1$ (see Figure~\ref{fig:ecdfExample}). To extract a feature vector $\bm{f} \in \mathbb{R}^k$ from the distribution, we first divide the range of $\mathcal{P}_c$, i.e., $[0,1]$, into $k$ equally sized bins with centers respectively at $[b_1,b_2,\ldots,b_k]$. The $i^{th}$ feature component $f_i \in \mathbb{R}$ is then computed as:
\begin{eqnarray}
f_i = \mathcal{P}_c^{-1}(b_i)
\end{eqnarray}
Thus the feature vector $\bm{f}$ accurately captures the shape and positions from the underlying probability function $p(r)$, while the ECDF $\mathcal{P}_c$ can be computed efficiently using Kaplan-Meier estimator~\cite{cox84analysis}.  For completeness, Figure~\ref{fig:ecdfFeatures} shows the extracted ECDF-based feature vector  for $k=75$. The only parameter for the ECDF-based feature extraction method is the number of bins $k$, which controls the granularity with which the shape of the underlying distribution is captured. In our experiments we also append the mean of ratings in $R$ as a feature value to the extracted ECDF feature vector.

%\noindent {\it Adversarial Implications:} One problem of relying on all outgoing links is that a malicious website can hide by embedding a large number of web links to web pages with high ratings. To defend against such an attack while constructing the set $R$ (see above), we  only allow ratings $r \le C_r$, where $C_r$ is the {\em critical rating threshold} that helps to remove artificial inflation of the set $R$ with good reputation ratings. The choice of $C_r$ can be application dependent and ideally should be adapted based on the overall costs of making false negative predictions. 

\noindent {\it Adversarial Implications:} If ECDF features were based
on all outgoing links, a malicious website may attempt to evade
detection by embedding a large number of links to pages with high
ratings. To deter such an attack,
while constructing the set $R$ (see above),
we only allow ratings $r \le C_r$, where $C_r$ is the {\em critical
rating threshold}.
%This helps to remove artificial inflation of the set $R$ with good reputation ratings.
The choice of $C_r$ can be application dependent and ideally should be
adapted based on the overall costs of making false negative
predictions.

%ECDF-based features provide an unique opportunity to get important clues about a unrated web page, based on the user ratings of the web pages it refers to. We hypothesize that often a web page that links to other pages with poor ratings, is most likely going to be poor itself and we validate this hypothesis by conducting extensive set of experiments. 

%In our evaluation section we will study the influence of $C_r$ on prediction performance in detail.

\subsubsection{Topic Model-based Features}
\label{sssec:topic}

To gain further insight into the type of content on a web page, we analyze the text in the page and extract a set of features that captures the distribution over a set of topics. A topic is defined as a probability distribution over a fixed set of words. In order to learn the topics in an unsupervised manner, we employ the well established {\em Latent Dirichlet Allocation} (LDA) model~\cite{blei03latent}. The main objective of LDA, or in general in any topic modeling algorithm, is to extract short descriptions of documents, while preserving statistical relationships that are useful, e.g., for document summarization and classification. In this work, we only focus on text in English. As a significant portion of the webpages in our evaluation dataset (see Section~\ref{ssec:datasets}) is non-english we use Google translation APIs, as part of the web crawler, to convert text into english. To avoid translation errors, we use an english dictionary to validate words before they are included in the {\em vocabulary} set $V$ used by the LDA model.

The main objective of the LDA model is to learn model parameters, such as $K$ topics $\beta_{1:K}$, the topic proportions $\theta_d$ in the document $d$, and topic assignments $z_{n,d}$ of observed word $w_n$ in document $d$ from the corpus of webpages. A brief overview of the topic model and the definitions of the parameters are given in Appendix~\ref{app:topic}. Once the LDA parameters are learned, given the set of words $w$ present on a webpage and the topics $\beta_{1:K}$, the topic model-based feature set for the webpage is computed as: $p(\theta_d | w, \beta_{1:K})$, i.e., the estimated topic proportions.

\noindent {\em Adversarial Implications:} Similarly to the ECDF-based features, the topic model-based features can be exploited by an adversary. As the topic proportion term $p(\theta_d | w, \beta_{1:K})$ captures the relative weight of various topics being described within the text $w$, an attacker can simply add random words that can boost the probability of certain topics. In Section~\ref{sec:discussion} we propose a possible solution to prevent this attack.

\subsection{Ensemble Classification}
\label{ssec:classifier}

One challenge in the feature extraction procedures, described above, is that often one or more feature types are missing from a web page. For example, in reality, not all web pages use JavaScript, contain embedded forward links, or use textual descriptions, although the HTML features are always available. Thus, a new classification technique is required that is able to overcome the problem of feature unavailability. Existing approaches such as~\cite{canali11prophiler, Ma09BeyondBlacklists, seifert08identification, seifert08identify}, do not address this problem and therefore have limited generalizability. 

According to Bayesian theory~\cite{kittler98combining}, given HTML ($\bm{f}_H$), JavaScript ($\bm{f}_J$), ECDF ($\bm{f}_E$), and Topic ($\bm{f}_T$) feature vectors, a URL should be assigned to the class $c_j \in \{bad,\, good\}$, if the posterior probability for class $c_j$ is maximum, i.e.
\begin{eqnarray} 
assign \,\,\,\,\, URL &\rightarrow& c_j \,\,\,\, if \nonumber\\
p(c_j | \bm{f}_H,\bm{f}_J,\bm{f}_E,\bm{f}_T) &=& \max_{i} \,\,p(c_i | \bm{f}_H,\bm{f}_J,\bm{f}_E,\bm{f}_T)
\label{eqn:decisionRule}
\end{eqnarray} 

The computation of $p(c_j | \bm{f}_H,\bm{f}_J,\bm{f}_E,\bm{f}_T)$ depends on the joint probability functions (likelihood) $p(\bm{f}_H,\bm{f}_J,\bm{f}_E,\bm{f}_T|c_j)$ and the prior probability $p(c_j)$, i.e.:
\begin{eqnarray} 
p(c_j | \bm{f}_H,\bm{f}_J,\bm{f}_E,\bm{f}_T) \propto p(\bm{f}_H,\bm{f}_J,\bm{f}_E,\bm{f}_T | c_j) \, p(c_j)
\label{eqn:bayes}
\end{eqnarray} 
The likelihoods are difficult to infer when one or more features are unavailable. The likelihood computation can be simplified by combining decision support of individual classifiers on different feature types~\cite{kittler98combining}. Accordingly, we train four classifiers $\mathcal{C}_H$, $\mathcal{C}_J$, $\mathcal{C}_E$, and $\mathcal{C}_T$ using valid $\bm{f}_H$, $\bm{f}_J$, $\bm{f}_E$, and $\bm{f}_T$ features respectively, where each classifiers returns a posterior probability distribution over the {\em bad} and {\em good} classes. However during prediction, if a feature type is unavailable, we do not include the corresponding classifier while computing the overall posterior probabilities.

%While predicting the safety of a web page, each classifier returns the posterior probability distribution over the classes {\em bad} and {\em good}. the main objective is to predict the safety or class of a web page given its features $(\bm{f}_h, \bm{f}_j, \bm{f}_e, \bm{f}_t)$.
%Note that one or more of $(\bm{f}_j, \bm{f}_e, \bm{f}_t)$ can be missing. Given the classifier $\mathcal{C}_h$, we compute the posterior probabilities for all the target classes, i.e., we compute:
%\begin{eqnarray}
%p_h(c_i | \bm{f}_h), \forall i \in \{1,\ldots, C\},
%\end{eqnarray}
%where $C$ is the total number of classes present. Similarly, we compute the posterior probabilities $p_j(\cdot), p_e(\cdot)$ and $p_t(\cdot)$ using the other three classifiers. 

\begin{figure}[!t]
\centering
\includegraphics[width=0.95\linewidth]{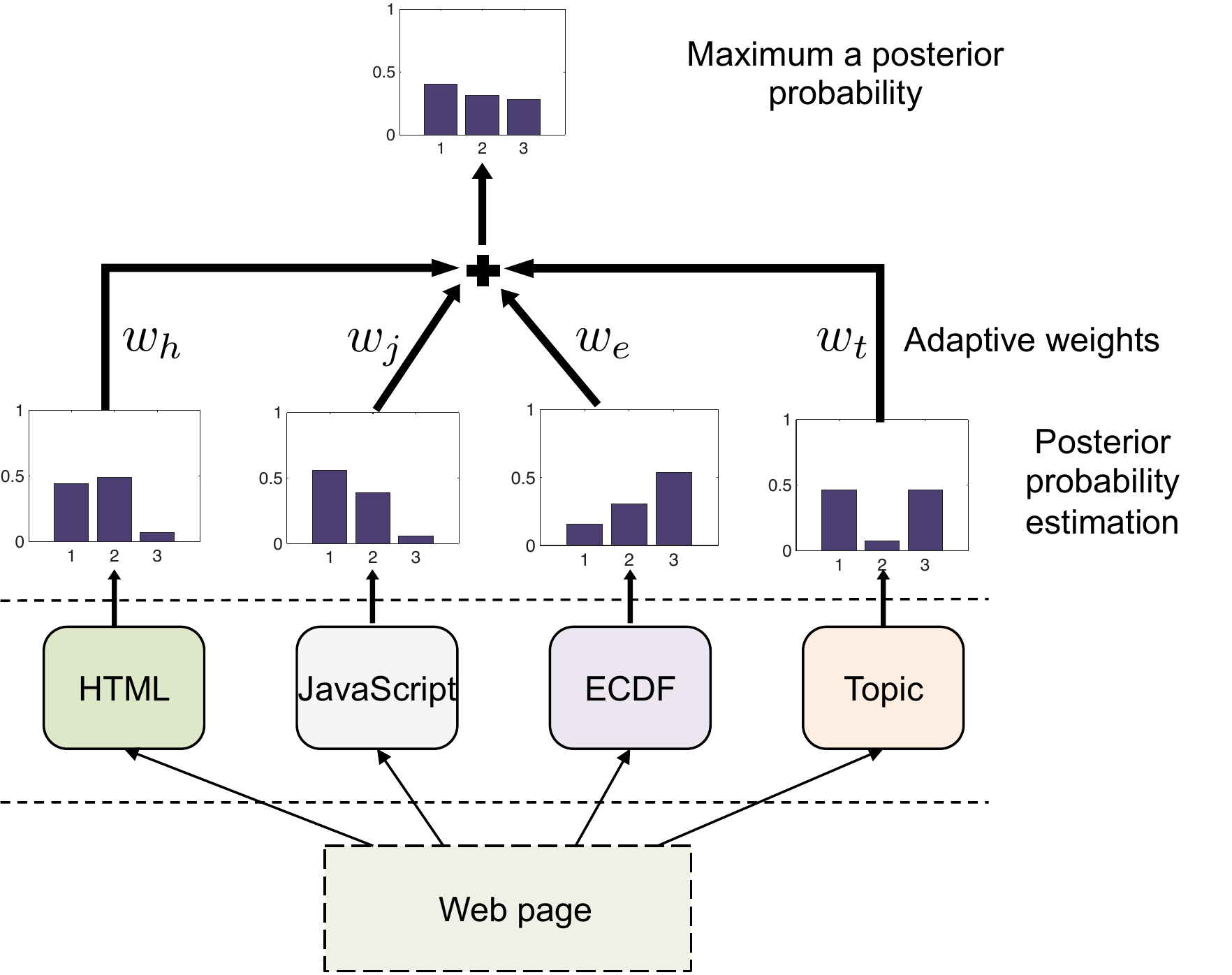}
\vspace{-3mm}
\caption{Overview of the ensemble classification approach used by \sys{}.}
\label{fig:ensembleClassifier}
\end{figure}

A number of strategies can be adopted to combine the posterior probabilities of the classifiers to generate the overall belief. In this paper we propose a linear combination rule that determines the combination weights of individual classifiers using the Fukunaga class separability score~\cite{fukunaga90introduction}. Our adaptive weight selection method is based on the intuition that a classifier should be given more importance if it is easy to separate among the {\em bad} and {\em good} classes in the corresponding feature space. See Appendix~\ref{app:fukunaka} for definition of class separability used in this work and other popular combination rules. For each classifier, we compute the separability score after correlation based feature subset selection. The separability scores, after normalization, are then used as the respective weight $w_k$ for the classifier $\mathcal{C}_k$. The final belief of the class $c_j$ is estimated as:
\begin{eqnarray} 
p^*(c_j | \bm{f}_H, \bm{f}_J, \bm{f}_E, \bm{f}_T) = \sum_{k \in \{H,J,E,T\}} w_k \,\, p(c_j | \bm{f}_k)
\label{eqn:map} 
\end{eqnarray}
The final predicted class $c_j$ is inferred by applying the decision rule given in Equation~\ref{eqn:decisionRule} using the computed belief above. 
Fig.~\ref{fig:ensembleClassifier} shows the data adaptive ensemble classification technique used by \sys{}. 
%\begin{eqnarray} 
%\hat{y} = arg \max_{j} \,\,p^*(c_j |  \bm{f}_H, \bm{f}_J, \bm{f}_E, \bm{f}_T)
%\end{eqnarray}

\section{Experimental Settings}
\label{sec:experiments}
%In this section we systematically report an extensive set of experiments conducted using real-world web pages.

\subsection{Datasets}
\label{ssec:datasets}

\begin{figure}[!t]
\centering
\includegraphics[width=0.9\linewidth]{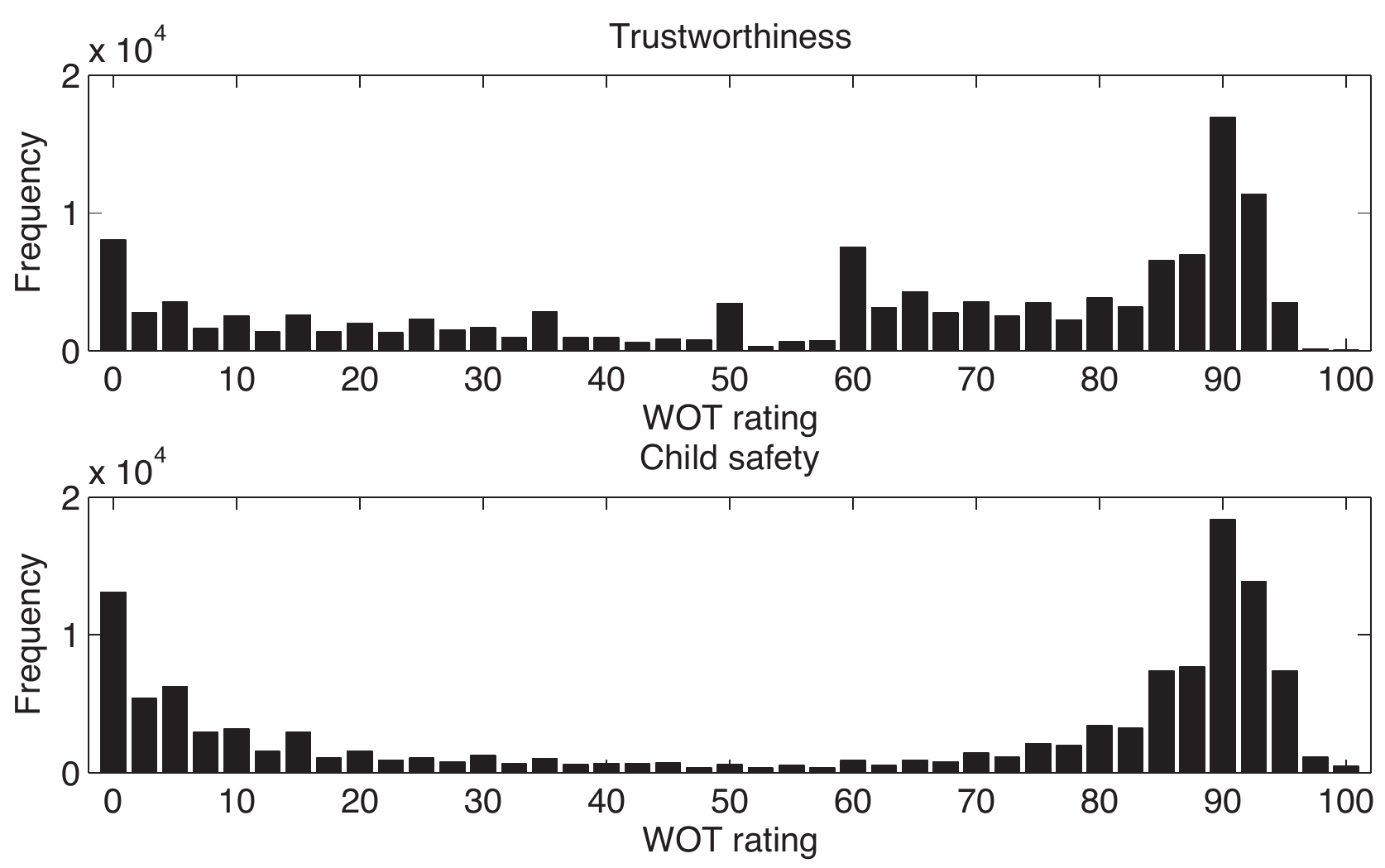}
\vspace{-3mm}
\caption{Histograms of all webpages in our dataset in two reputation dimensions: trustworthiness (top) and child safety (bottom).}
\label{fig:histDataset}
\end{figure}

To perform an extensive and systematic study, we generated a pool of over $140,000$ URLs and obtained their reputation ratings in both dimensions using the WOT API. Out of these, $80,000$ URLs have positive reputation ratings, and $60,000$ URLs have negative ratings. For each URL we crawl the web page to extract HTML, JavaScript, ECDF and topic model features where available. Figure~\ref{fig:histDataset} illustrates the histograms of reputation ratings for all webpages in our dataset.
The dataset, where at least valid HTML  features and WOT ratings are available, is referred to as the {\em opportunistic} dataset. 
Out of $140,000$ URLs, $89,220$ web pages
have trustworthiness ratings, and $84,714$ web pages have ratings for child safety.
However, the number drops to $31,995$, in case of trustworthiness, and to $38,118$ for child safety, when validity of all feature types are considered (for $\bm{\mathcal{T}_h} = 40 $). We refer to this second dataset as the {\em all-valid} dataset. The significant drop in the size of the all-valid dataset further highlights that feature unavailability is intrinsic to web data analysis. 

%\subsubsection*{Malware Dataset}
%\label{sssec:malwareData}

Existing research primarily focussed on detecting if a webpage is malicious. However, the malware dataset used in~\cite{canali11prophiler} is no longer available, which makes exact replication of Prophiler results difficult. The definition of trustworthiness does not directly correspond with malware.
In addition to the reputation ratings, WOT provides category information, such as `malware', `scam',  `suspicious' and `good site', of websites based on votes from users and third parties. We filter the all-valid dataset and generate a malware dataset consisting of $2,784$ webpages that were categorized by WOT as `malware or virus' and contain all feature types. To generate a dataset containing both malware and benign webpages, we include an equal number of webpages that got very high trustworthiness ratings and have all feature types. We refer to this dataset as the {\em malware} dataset. 
%Fo example, Figure~\ref{fig:malware} depicts the histogram of webpage trustworthiness ratings of our malware dataset.
%, although WOT expects that majority of malware URLs would receive low trustworthiness ratings. 

%\begin{figure}[!t]
%\centering
%\includegraphics[width=\linewidth]{malwareClassDistribution}
%\vspace{-7mm}
%\caption{Histograms showing trustworthiness ratings of webpages in our {\em malware} dataset.}
%\label{fig:malware}
%\end{figure}

%\subsubsection*{Two-category Dataset}
%\label{sssec:twocategory}

Lastly, we construct another dataset by considering only the URLs that fall either in the top most or the bottom most trustworthiness rating categories, see Figure~\ref{fig:wot_score} for definitions of various rating categories used by WOT. As with malware dataset, we only consider webpages for which all feature types are available. Our {\em two-category} dataset consists of $10,118$ very poor and $13,539$ excellent webpages.

\subsection{Baseline Algorithms}
\label{ssec:prophiler}

In our experiments, we report comparison results against Prophier~\cite{canali11prophiler}. Prophiler relies on HTML, JavaScript, and URL/HOST features to detect if a webpage is malicious. However, it uses APIs to a proprietary WHOIS \cite{whois} system
and uses a private database for blacklisted URLs to extract URL/HOST features.
Neither of these are available openly, which makes the corresponding URL/HOST feature vectors invalid for our datasets.
%API inaccessibility and unavailability of suitable blacklisted database covering WOT URLs used in our dataset makes the majority of the URL/HOST feature vectors in our experiments invalid.
Consequently, we do not use URL/HOST features in our ensemble classification system.

Note that it is very easy to incorporate additional feature types to our classification system, e.g., training a classifier $\mathcal{C}$ using the new feature type and then considering the posterior probabilities in Equation~\ref{eqn:map}. Contrary to our approach, i.e., assigning data driven weighting of classifiers to compute the final belief (see Section~\ref{ssec:classifier}), Prophiler uses the `OR' combination rule (see Appendix~\ref{app:rules}).
We systematically compare the performance of \sys{} with the ensemble classification methodology considering different subsets of feature types.

%To gain insights on how different feature types influence web safety detection results, we consider several versions of the \sys{} system, where we systematically vary the feature sets used for web safety detection.

\subsection{Evaluation Metric}
\label{ssec:metric}

We use $10$-fold cross validations when presenting classification performance for all the approaches. As the primary performance metric, we use Avg. F$_1$-score, FNR, and FPR. The definitions of all the evaluation metrics are given in Appendix~\ref{app:fpr}.

\section{Evaluation}
\label{sec:eval}

We begin our evaluation by first considering classification performance on the all-valid dataset using Random Forest as the basic classifier\footnote{We also experimented using linear-SVM, SVM, KNN and C4.5 classifiers, and chose Random Forest for its superior performance.}. Note that, all URLs considered within this dataset have valid HTML (H), JavaScript (J), ECDF (E) and Topic-based (T) features. This dataset allows us to systematically study the influence of various feature combinations on the overall classification performance of \sys{}. Table~\ref{tbl:allvalidth40} summarizes the performance of \sys{} in both reputation dimensions with the parametric settings $\bm{\mathcal{T}_h} = 40$ (see Equation~\ref{eqn:classification}), and $C_r = 40$ (see Section~\ref{ssec:contentFeatures}). Additionally, the table also includes the performance of Prophiler.

\begin{table}[!t]
\centering
\begin{threeparttable}
\caption{Performance of \sys{} (under various feature combinations) and Prophiler on the all-valid dataset.}

\begin{tabular} {|c|c|c|c|c|c|c|}
\multicolumn{7}{l}{\bf All-valid dataset size: 31,995 URLs,  $\bm{\mathcal{T}_h} = 40$, $C_r = 40$}\\
\multicolumn{7}{l}{\bf Reputation dimension: Trustworthiness}\\
\hline
\multicolumn{4}{|c|}{\bf Feature sets}      &  {\bf Avg. F$_1$-Score} & {\bf FNR} & {\bf FPR}\\  
\cline{1-4}
{\bf H} & {\bf J} & {\bf E} & {\bf T} & (\%) & (\%) & (\%)\\
\hline
\checkmark &  &  &   & $75.6$ ** & $25.2$ & $23.7$\\
& \checkmark &  &  &  $74.3$ **& $25.6$ & $25.9$\\
& & \checkmark &  &  $66.9$ **& $33.5$ & $32.7$\\
&  & & \checkmark &  $74.5$ **& $26.8$ & $24.3$\\
\hline
\checkmark & \checkmark &  &  &  $76.9$ **& $23.3$ & $22.8$\\
\checkmark &  & \checkmark &  &  $77.3$ **& $23.7$ & $21.7$\\
\checkmark &  &  & \checkmark &  $78.5$ **& $21.8$ & $21.3$\\
& \checkmark & \checkmark &  &  $72.1$ **& $28.9$ & $26.9$\\
& \checkmark &  & \checkmark &  $77.1$ **& $24.0$ & $21.9$\\
&  & \checkmark & \checkmark &  $77.5$ **& $23.9$ & $21.2$\\
\hline
\checkmark & \checkmark &  \checkmark &  &  $78.8$ **& $21.6$ & $20.7$\\
\checkmark &  \checkmark &  & \checkmark &  $79.5$ **& $20.9$ & $20.1$\\
\checkmark &  & \checkmark & \checkmark &  $80.4$ **& $20.2$ & $19.0$\\
&  \checkmark & \checkmark & \checkmark &  $79.6$ **& $21.4$ & $19.4$\\
\hline
\rowcolor{Gray}
\checkmark & \checkmark & \checkmark & \checkmark &  \bm{$81.3$} & \bm{$19.0$} & \bm{$18.3$}\\

\hline
\multicolumn{7}{l}{}\\
\hline
\multicolumn{4}{|c|}{\bf Prophiler} &  $74.5$ **& $14.2$ & $35.9$\\
\hline
\multicolumn{7}{l}{}\\
\multicolumn{7}{l}{\bf All-valid dataset size: 38,118 URLs,  $\bm{\mathcal{T}_h} = 40$, $C_r = 40$}\\
\multicolumn{7}{l}{\bf Reputation dimension: Child safety}\\
\hline
\multicolumn{4}{|c|}{\bf Feature sets}      &  {\bf Avg. F$_1$-Score} & {\bf FNR} & {\bf FPR}\\  
\cline{1-4}
{\bf H} & {\bf J} & {\bf E} & {\bf T} & (\%) & (\%) & (\%)\\
\hline
 \checkmark &  &  &   & $80.3$ **& $15.2$ & $25.8$\\
& \checkmark &  &  &  $79.6$ **& $15.8$ & $26.8$\\
& & \checkmark &  &  $73.1$ **& $22.0$ & $33.6$\\
&  & & \checkmark &  $81.9$ **& $17.0$ & $19.8$\\
\hline
\checkmark & \checkmark &  &  &  $81.1$ **& $14.3$ & $25.1$\\
\checkmark &  & \checkmark &  &  $79.4$ **& $16.3$ & $26.3$\\
\checkmark &  &  & \checkmark &  $84.5$ **& $13.4$ & $18.3$\\
& \checkmark & \checkmark &  &  $77.3$ **& $18.1$ & $29.1$\\
& \checkmark &  & \checkmark &  $83.9$ **& $14.6$ & $18.2$\\
&  & \checkmark & \checkmark &  $83.9$ **& $14.9$ & $17.7$\\
\hline
\checkmark & \checkmark &  \checkmark &  &  $82.4$ **& $13.5$ & $23.2$\\
\checkmark &  \checkmark &  & \checkmark &  $85.2$ **& $12.2$ & $18.4$\\
\checkmark &  & \checkmark & \checkmark &  $85.7$ **& $12.6$ & $16.7$\\
&  \checkmark & \checkmark & \checkmark &  $85.3$ **& $13.4$ & $16.6$\\
\hline
\rowcolor{Gray}
\checkmark & \checkmark & \checkmark & \checkmark &  \bm{$86.4$} & \bm{$11.6$} & \bm{$16.2$}\\
\hline
\multicolumn{7}{l}{}\\
\hline
\multicolumn{4}{|c|}{\bf Prophiler} &  $79.5$ **& $9.6$ & $34.5$\\
\hline
\end{tabular}
\label{tbl:allvalidth40}
\begin{tablenotes}
\footnotesize
\item **: Statistically significant with $99\%$ confidence.
%\item * : Statistically significant with $95\%$ confidence.
\end{tablenotes}
\end{threeparttable}
\vspace{-3mm}
\end{table}

For trustworthiness, \sys{} achieves the highest Avg.\ F$_1$-score of $81.3\%$, when all feature types are considered (highlighted in gray), at the same time achieving the lowest FNR ($19\%$) and FPR ($18.3\%$). Similarly for child safety, \sys{} with all feature types achieves the best performance ($86.4\%$), lowest FNR ($11.6\%$) and lowest FPR ($16.2\%$). In both reputation dimensions, the performance using all features, is significantly better (statistically) than all other feature combinations, i.e., $p \ll 0.01$ in McNemar $\bm{\chi}^2$ test with Yates' correction~\cite{mcnemar47note}. 

Prophiler shows a statistically weaker classification performance in both reputation dimensions compared to \sys{} (employing all feature types). However, it achieves a better FNR in prediction than \sys{}. This is due to the use of a conservative `OR'  classifier combination rule (see Appendix~\ref{app:rules}) that is more likely to report a URL as bad. This higher likelihood of predicting webpages as bad improves the overall recall of the bad class, which consequently pulls down the FNR for Prophiler, however, at the expense of a higher FPR. Prophiler focuses solely on reducing FNR. In contrast, in use cases where overall usability in prediction is important, both FNR and FPR should be reduced. For example, in predicting safety ratings, a low FPR is also needed to avoid showing frequent warnings to users for actually good websites. A very similar classification performance (Table~\ref{tbl:allvalidth60} in the Appendix) is observed when the same set of experiments are conducted on the all-valid dataset for $\bm{\mathcal{T}_h} = 60$, and $C_r = 60$.

\begin{table}[!t]
\centering
\begin{threeparttable}
\caption{Performance of \sys{} on the opportunistic dataset under various classifier combination rules.}

\begin{tabular} {|c|c|c|c|c|}
\multicolumn{5}{l}{\bf Opportunistic dataset size: 89,220 URLs, $\bm{\mathcal{T}_h} = 40$ and $C_r=40$}\\
\multicolumn{5}{l}{\bf Reputation dimension: Trustworthiness}\\
\hline
\multicolumn{2}{|c|}{\bf Experiment}      &  {\bf Avg. F$_1$-Score} & {\bf FNR} & {\bf FPR}\\  
\cline{1-4}
%$\bm{\mathcal{T}_h}$ & $C_r$ & {\bf Balancing} & {\bf Comb. Rule} & (\%) & (\%) & (\%)\\
{\bf Comb. Rule} & {\bf Balancing} & (\%) & (\%) & (\%)\\
\hline
\rowcolor{Gray}
{\bf Adaptive} & & $78.0$ & $56.4$ & $4.1$\\
Sum & & $77.8$ & $57.9$ & \bm{$3.5$}\\
Product & & \bm{$78.9$} **& $53.3$ & $4.6$\\
Or & & $32.4$ ** & \bm{$10.1$} & $84.2$\\
Voting & & $71.4$ **& $38.8$ & $25.2$\\
\hline
Prophiler* & & $72.6$ **& $45.3$ & $19.8$\\
\hline
\hline
\rowcolor{Gray}
{\bf Adaptive} & \checkmark & \bm{$74.0$} & $22.3$ & $29.1$\\
Sum & \checkmark & $74.0$ & $22.3$ & \bm{$29.0$}\\
Product & \checkmark & $73.6$ **& $22.8$ & $29.5$\\
Or & \checkmark & $27.3$ **& \bm{$2.3$} & $89.8$\\
Voting & \checkmark & $57.3$ **& $11.1$ & $57.0$\\
\hline
Prophiler* & \checkmark & $62.0$ **& $14.4$ & $49.8$\\
\hline
\multicolumn{5}{l}{}\\
\multicolumn{5}{l}{\bf Opportunistic dataset size: 84,714 URLs, $\bm{\mathcal{T}_h} = 40$ and $C_r=40$}\\
\multicolumn{5}{l}{\bf Reputation dimension: Child safety}\\
\hline
\multicolumn{2}{|c|}{\bf Experiment}      &  {\bf Avg. F$_1$-Score} & {\bf FNR} & {\bf FPR}\\  
\cline{1-4}
{\bf Comb. Rule}  & {\bf Balancing} & (\%) & (\%) & (\%)\\
\hline
\rowcolor{Gray}
{\bf Adaptive} & & \bm{$83.7$} & $29.8$ & $7.2$\\
Sum & & $83.4$ \mbox{ *} & $31.2$ & \bm{$6.6$}\\
Product & & $83.6$ & $29.6$ & $7.5$\\
Or & & $40.7$ **& \bm{$4.7$} & $82.3$\\
Voting & & $73.9$ **& $19.7$ & $30.7$\\
\hline
Prophiler* & & $73.9$ **& $26.7$ & $26.2$\\
\hline
\hline
\rowcolor{Gray}
{\bf Adaptive} & \checkmark &  \bm{$81.5$} & $25.3$ & \bm{$14.0$}\\
Sum & \checkmark & $81.1$ **& $22.9$ & $16.4$\\
Product & \checkmark & $80.9$ **& $23.1$ & $16.8$\\
Or & \checkmark & $40.0$ **& \bm{$3.0$} & $83.4$\\
Voting & \checkmark & $68.1$ **& $12.3$ & $44.3$\\
\hline
Prophiler* & \checkmark & $69.4$ **& $15.6$ & $40.5$\\
\hline
\end{tabular}
\label{tbl:opportunisticth40}
\begin{tablenotes}
\footnotesize
\item **: Statistically significant with $99\%$ confidence.
\item * : Statistically significant with $95\%$ confidence.
\end{tablenotes}
\end{threeparttable}
\vspace{-3mm}
\end{table}

In reality, not all feature types are available for all URLs. To evaluate the performance of \sys{} under real life situations we next present results on the opportunistic dataset. In these experiments, we only present the performance of \sys{} while considering all available feature types. Moreover, we study the performance of various classifier combination rules and present the results in Table~\ref{tbl:opportunisticth40} for both reputation dimensions. In contrast to the all-valid dataset, $\bm{\mathcal{T}_h} = 40$, generates a high degree of class imbalance in our opportunistic dataset (see Figure~\ref{fig:histDataset}). During the training phase the prevalence of one class affects the process of learning, and the learned classifier is often biased towards the over-represented class~\cite{menardi14training}. To mitigate class imbalances during training, we also report experimental results when a simple class balancing approach, i.e., reducing data from the prevalent class, is applied during classifier training. The data driven, adaptive classification combination rule of \sys{} generates the best classification performance, with a notable exception in the case of unbalanced dataset for trustworthiness, where the `Product' rule achieves the highest Avg. F$_1$-Score.

%\begin{table}[!t]
%\caption{Confusion Matrix and Overall Performance of Prophiler on the malware dataset.}
%\begin{tabular}{l|l|c|c|ccc}
%\multicolumn{2}{c}{}&\multicolumn{2}{c}{\bf Predicted}& & &\\
%\cline{3-4}
%\multicolumn{2}{c|}{}& {\bf Malware} & {\bf Benign} &\multicolumn{1}{c}{\bf Total} & {\bf Avg. F$_1$-Score:} & {\bf 80.7}\%\\
%\cline{2-4}
%\multirow{2}{*}{\bf Actual}& {\bf Malware} & $2475$ & $309$ & $2784$ & {\bf FNR:} & {\bf 11.1}\%\\
%\cline{2-4}
%& {\bf Benign} & $761$ & $2023$ & $2784$ & {\bf FPR:} & {\bf 27.3}\%\\
%\cline{2-4}
%\multicolumn{1}{c}{} & \multicolumn{1}{c}{\bf Total} & \multicolumn{1}{c}{$3236$} & \multicolumn{1}{c}{$2332$} & \multicolumn{1}{c}{\bf 5568} & & \\
%\end{tabular}
%\label{tbl:proMalware}
%\end{table}
%
%\begin{table}[!t]
%\caption{Confusion Matrix and Overall Performance of \sys on the malware dataset.}
%\begin{tabular}{l|l|c|c|ccc}
%\multicolumn{2}{c}{}&\multicolumn{2}{c}{\bf Predicted}& & &\\
%\cline{3-4}
%\multicolumn{2}{c|}{}& {\bf Malware} & {\bf Benign} &\multicolumn{1}{c}{\bf Total} & {\bf Avg. F$_1$-Score:} & {\bf 89.0}\%\\
%\cline{2-4}
%\multirow{2}{*}{\bf Actual}& {\bf Malware} & $2497$ & $287$ & $2784$ & {\bf FNR:} & {\bf 10.3}\%\\
%\cline{2-4}
%& {\bf Benign} & $324$ & $2460$ & $2784$ & {\bf FPR:} & {\bf 11.6}\%\\
%\cline{2-4}
%\multicolumn{1}{c}{} & \multicolumn{1}{c}{\bf Total} & \multicolumn{1}{c}{$2821$} & \multicolumn{1}{c}{$2747$} & \multicolumn{1}{c}{\bf 5568} & & \\
%\end{tabular}
%\label{tbl:laMalware}
%\end{table}

\begin{table}[!t]
\centering
\begin{threeparttable}
\caption{Performance of \sys{} and Prophiler on the malware and two-category datasets.}

\begin{tabular} {|c|c|c|c|c|c|c|}
\multicolumn{7}{l}{\bf Malware dataset size: 5,568 URLs}\\
\hline
\multicolumn{4}{|c|}{\bf Feature sets}      &  {\bf Avg. F$_1$-Score} & {\bf FNR} & {\bf FPR}\\  
\cline{1-4}
{\bf H} & {\bf J} & {\bf E} & {\bf T} & (\%) & (\%) & (\%)\\
\hline
\rowcolor{Gray}
\checkmark & \checkmark & \checkmark & \checkmark &  \bm{$89.0$} & \bm{$10.3$} & \bm{$11.6$}\\

\hline
\multicolumn{7}{l}{}\\
\hline
\multicolumn{4}{|c|}{\bf Prophiler} &  $80.7$ **& $11.1$ & $27.3$\\
\hline
\multicolumn{7}{l}{}\\
\multicolumn{7}{l}{\bf Tow-category dataset size: 23,657 URLs}\\
\hline
\multicolumn{4}{|c|}{\bf Feature sets}      &  {\bf Avg. F$_1$-Score} & {\bf FNR} & {\bf FPR}\\  
\cline{1-4}
{\bf H} & {\bf J} & {\bf E} & {\bf T} & (\%) & (\%) & (\%)\\
\hline
\rowcolor{Gray}
\checkmark & \checkmark & \checkmark & \checkmark &  \bm{$89.8$} & \bm{$13.9$} & \bm{$7.4$}\\
\hline
\multicolumn{7}{l}{}\\
\hline
\multicolumn{4}{|c|}{\bf Prophiler} &  $79.3$ **& $16.2$ & $24.3$\\
\hline
\end{tabular}
\label{tbl:maltwocat}
\begin{tablenotes}
\footnotesize
\item **: Statistically significant with $99\%$ confidence.
%\item * : Statistically significant with $95\%$ confidence.
\end{tablenotes}
\end{threeparttable}
\vspace{-3mm}
\end{table}

Prophiler has previously been shown to perform well in detecting malicious websites. To show how \sys{} (with all features) perform in such scenarios, we repeated the experiments on the malware and two-category datasets and present the results in Table~\ref{tbl:maltwocat}. \sys{} achieves average F$_1$-scores of $89\%$ for the malware dataset and $89.8\%$ for the two-category dataset, which are significantly better ($p \ll 0.01$) than Prophiler's performance of $80.7\%$ for the malware dataset and $79.3\%$ for the two-category dataset.
\sys{} also generates better FNR and FPR than Prophiler in both cases.

\section{Discussion}
\label{sec:discussion}

%\noindent\textbf{Feature importances:}
\subsection{Feature Importance in Reputation Prediction}

\begin{figure}[!t]
\centering
\includegraphics[width=0.9\linewidth]{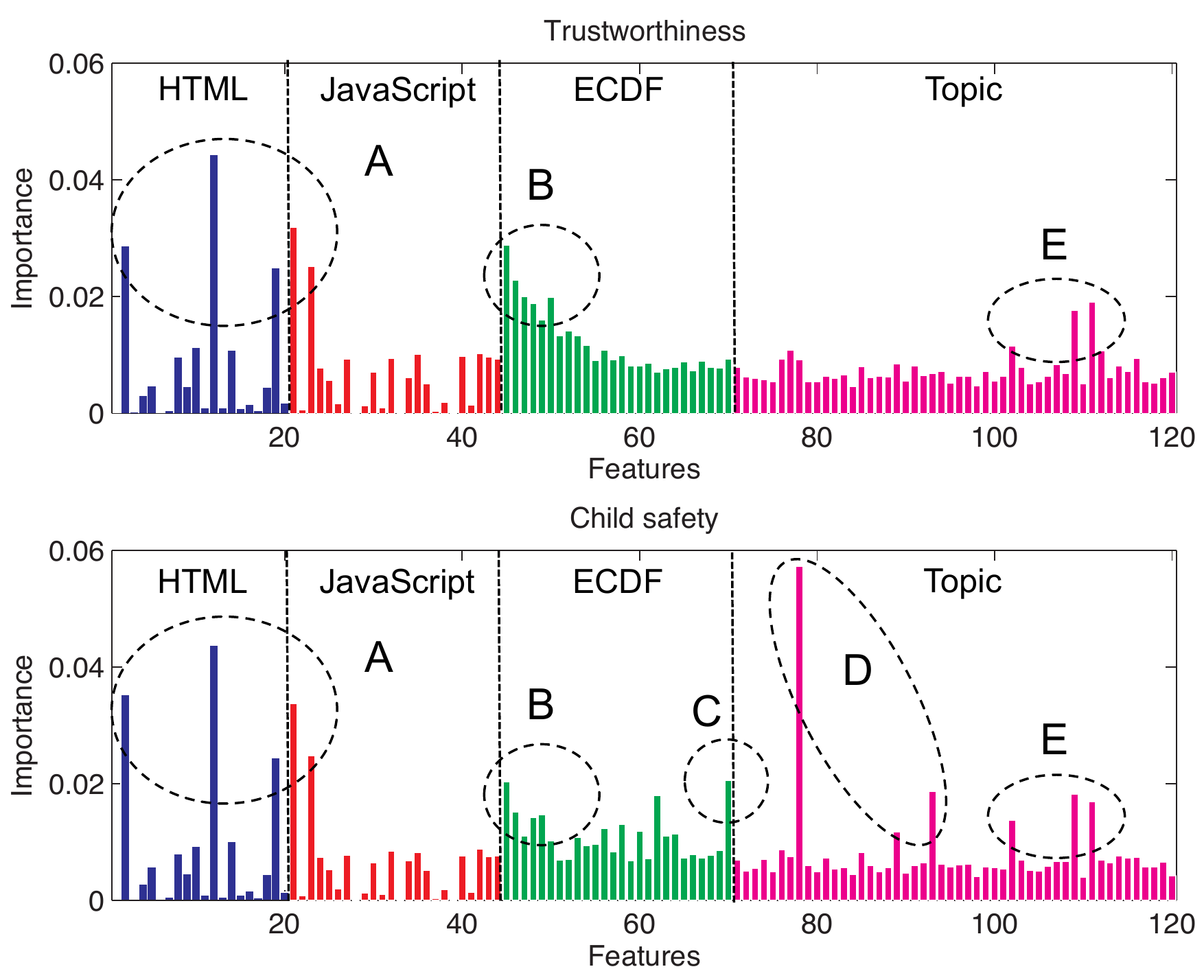}
\vspace{-3mm}
\caption{Importance of individual features, while predicting trustworthiness and child safety, computed by the Random Forest classifier on the all-valid dataset.}
%Feature importances. Each bar represents the importance of a specific feature of either HTML (blue), Javascript (Red), ECDF (Green) or Topic (Black) categories.}
\label{fig:featureimportance}
\vspace{-3mm}
\end{figure}

Our results show that the structural and content related properties of a website can be effectively used to predict not only its maliciousness, but also the more challenging properties of trustworthiness and child safety. In order to understand the overall classification results, we study the importance\footnote{Feature importance is defined as the total decrease in node impurity averaged over all the trees~\cite{breiman2001random}.} of individual features as computed by a Random Forest classifier. In Fig.~\ref{fig:featureimportance} we plot the average\footnote{Over $10$ folds.} importance for all (120) features used in this work when training a Random Forest classifier (using $100$ trees) on the all-valid dataset ($\bm{\mathcal{T}_h}=40, C_r=40$). The higher the value, the more important is the feature. Fig.~\ref{fig:featureimportance} further highlights that different features are assigned different relative importances while separating good websites from bad ones in each reputation dimension.

%From this we can see which properties of websites best separate good websites from bad ones in each dimension.

%{\color{red}Positive feature importance values correlate with bad websites whereas negative values correlate with good ones.}

Interestingly, the importance scores of the HTML and JavaScript-based features look very similar for both trustworthiness and child safety predictions. The most important features, shown by the dotted region A in the figure, are related to script tags in HTML, direct assignments in JavaScript, and the total character count in both. Although a few structural (i.e., HTML and JavaScript) features are found to be important, a majority of them have little or no significance. Contrary to the structural features, ECDF features show significant differences in importance scores for the two reputation dimensions. For trustworthiness, low ratings of the embedded forward links (region B) play an important role in prediction. In child safety, the mean value of the embedded ratings (region C) plays a significant role also. For trustworthiness, the three most important topics (region E) are related to money-making, news, and weather. Among the rest of the topic features, none are significantly better or worse than the others. For child safety prediction there are three other topics (region D) that play a significant role and as expected, these topics correspond to adult content.

Although, we use the same feature set for predicting both reputation dimensions, the feature selection inherent to the Random Forest classifier learns very different mapping functions for each prediction task. Fig.~\ref{fig:featureimportance} provides evidence that our proposed ECDF and Topic-based features contribute consistently in predicting subjective ratings of web pages.  

%emphasizes that the a very few structural features contributes to the prediction tasks, whereas majority of the content-based features 

%\noindent\textbf{Here the questions are just to help me structure the Section /Otto}:

%\noindent\textbf{STRUCTURE: Why Prophiler performs badly on TW/CSafety?}:

%\subsection{Classification performance}

%We find that LookAhead outperforms Prophiler, which considers only structural features of a websites. This is consistent with what we observe based on feature importances (Fig.~\ref{fig:featureimportance}), and confirms the intuition that trustworthiness and child safety are reflected more in the content of a page than in its inherent structure.More specifically, while a few HMTL and Javascript features have a large impact on classification, a majority play little or no role in predicting website ratings either way.

%In addition, when we consider that two out of three of the most important HTML features are related to scripts, we can conclude that HTML features by themselves have limited applicability for predicting the rating of a website.

%Observing the feature importances also seems to confirm that our choice of ECDF and topic-based features can be successfully used to predict reputation ratings. These features contribute more consistently into the classification performance unlike structural features.

\iffalse
In addition, we can note that certain topic areas seem to be significantly better at predicting the Child Safety of a website than others. This indicates that, given that a site owner wants to promote these topics on their website, it will be relatively easy to flag that website as unsafe for children.
\fi

%It is not clear how much this is the result
%of e.g. untrustworthy websites sharing similar structural features and 
%how much is due to popular opinion being swayed by certain web design %decisions. For example, the use of pop-ups on any kind of site may result %in a worse rating regardless of the trustworthiness of a website.

%\noindent\textbf{CONTENT: Why LookAhead performs like it does?}:
%For LookAhead, we can see that considering also the content-based features improves classification performance and so the FPR and TPR rates are reduced significantly.
%Comparing feature importances between structural and content-based features, we see already from Figure Fig.~\ref{fig:featureimportance} that the latter contribute more consistently into the overall classification. In addition we note that certain topics are better at predicting Child Safety than Trustworthiness, namely adult themed topics play a significant role when predicting the Child Safety of a webpage.

\subsection{Tuning of Prediction Performance}
Predictive performance of \sys{} can be primarily influenced by a number of factors: (i) the type of features considered (e.g., HTML and ECDF), (ii) the type of classifier used to train on individual feature dimensions (e.g., Random Forest and SVM),  (iii) strategies used to overcome class imbalances in the training data, and (iv) the combination rule used for computing the final posterior probability (e.g., Adaptive and Sum rule). Often, once the prediction pipeline is deployed, the factors (i)--(iii) are kept constant, as they are time consuming to re-build. However, the classifier combination strategy can be adapted in real time to control the overall performance of the \sys{} system. Based on the requirements, the system administrator can focus more on lowering the overall FNR by using the `OR' combination rule, e.g., while predicting child safety a very low FNR is expected for parental filtering systems. As evident from Table~\ref{tbl:opportunisticth40}, often emphasizing FNR inflates FPR. Our \sys{} system demonstrates a good balance of both FNR and FPR.

\subsection{Detection Rate}
When considering the implications of our results, in addition to the FNR and FPR, the proportion of good and bad websites in the wild should also be taken into account. In reality, this so-called base rate $B_r$, is biased towards good websites. Thus we look at the detection rate
for bad websites, i.e., what percentage of webpages that our classifier predicts as bad are truly bad.
 From WOT statistics~\cite{wotstat}, we see that roughly 20\% of websites that have a rating are dangerous regarding either trustworthiness or child safety. We use this number as our estimate for $B_r$, and compute the detection rate as:
\begin{eqnarray}
\mathcal{D}_r = \frac{(1 - FNR)\cdot B_r}{(1 - FNR)\cdot B_r + FPR\cdot(1 - B_r)},
\end{eqnarray}
%where FNR and FPR are false negative and positive rates, and BR is the base rate.
 
%for our previous classification scenarios with the FNR and FPR rates for both LookAhead and Prophiler. The detection rate is defined as:

\begin{table}[!t]
\centering
\caption{Detection rates for various classifiers settings.}
\begin{tabular} {|c|c|c|c|c|}
\hline
{\bf Dataset} & {\bf System} & {\bf FNR} & {\bf FPR} & {\bf $\mathcal{D}_r$}\\ 
& & (\%) & (\%) & (\%)\\
\hline
\rowcolor{Gray}
All-valid, TW & LookAhead & 19.0 & 18.3 & 52.5 \\
All-valid, TW & Prophiler & 14.2 & 35.9 & 37.4 \\
\hline
\rowcolor{Gray}
All-valid, CS & LookAhead & 11.6 & 16.2 & 57.5 \\
All-valid, CS & Prophiler & 9.6 & 34.5 & 39.6 \\
\rowcolor{Gray}
\hline
Opportunistic, TW & LookAhead & 22.3 & 29.1 & 40.0 \\
Opportunistic, TW & Prophiler & 14.4 & 49.8 & 30.1 \\
\hline
\rowcolor{Gray}
Opportunistic, CS & LookAhead & 25.3 & 14.0 & 57.2 \\
Opportunistic, CS & Prophiler & 15.6 & 40.5 & 34.3 \\
\hline
\end{tabular}
\label{tbl:detectionrates}
\vspace{-3mm}
\end{table}

Table~\ref{tbl:timePerformance} presents detection rates for our different classification scenarios. The datasets are presented for both trustworthiness (TW) and child safety (CS). We can see that due to the biased base rate, the detection rates are in the range of 30-40\% for Prophiler and 40-57\% for LookAhead. Still, the better classification performance of LookAhead over Prophiler is apparent here too. For example, if we consider a warning system for users, we can see that, in the case of all features being present, 52.5\% of possible warnings for untrustworthy webpages would be correct for LookAhead, with the rest being false alarms. The corresponding detection rate of 37.4\% for Prophiler is significantly lower.
This shows that while in general the problem of predicting a reputation rating is challenging due in part to the biased base rate, considering content-based features significantly increases the detection rate.

\subsection{Applications}

We see two potential uses for LookAhead:
\begin{itemize}
\item \emph{fast-tracking publication of ratings}: Crowdsourced
  reputation rating services like WOT do not announce a rating for a
  web site until they have enough input ratings to reach a sufficient
  level of confidence. If a partially accumulated rating (that has not
  reached a sufficient level of confidence) matches the rating
  predicted by our classifier, the reputation service may choose to
  fast-track the publication of the rating.
\item \emph{intermediate user feedback}: If a user attempts to
  navigate to an unrated page that is predicted by our classifier to
  have a potentially bad rating, the browser extension can warn the
  user accordingly.
\end{itemize}
Earlier research~\cite{tyler08evaluating} raised concerns about the
usefulness of crowdsourcing for security and privacy applications.
Nevertheless, given the popularity of systems like WOT, we argue that
a tool like LookAhead is essential for the security of users who have
chosen to rely on such systems. Also, note that although our analysis was done with WOT as the target rating system,
the methodology is applicable to any website safety rating system,
whether crowdsourced or expert-rated.  
%WOT ratings are for domains, rather than individual pages.  

\subsection{Performance considerations}

We summarize the performance of our various feature extraction techniques and report the average measured running time needed for computing them. For the purpose of computing the average extraction time we randomly selected $1,000$ URLs from our dataset and measured the time required to extract different classes of features on a standard Linux desktop computer (8 Gb RAM, 2.4 GHz processor) for the corresponding pages. In the case of Topic model features we also recorded the time for performing translation of non-english web pages. Table~\ref{tbl:timePerformance} summarizes the time analysis of our feature extraction methods.
The time of  $3.1 \,\,s$ that LookAhead needs for extracting structural features is comparable to that of  $3.06 \,\,s$ reported by Prophiler. When including the content based features, in total, LookAhead needs $6.3 \,\,s$ to extract all features from an English-language web page (and  $8.4 \,\,s$ if translation is needed).

%On average $3.1 \,\,s$ is needed per web page to extract the structural features, whereas only $1.9 \,\,s$ per web page is needed for obtaining the ECDF features. The topic model feature computation requires the least amount of average time per web page ($1.3 \,\,s$), when no translation is required. However, for non English web pages the average feature computation time increases to $3.4 \,\,s$ per page. The feature set collection time needed in our approach, especially for the content features (without translation), is smaller than the reported time of $3.06 \,\, s$ by Prophiler. Thus, our proposed feature extraction procedure is highly suitable for real time safety prediction of web pages.

\begin{table}[!t]
\centering
\caption{Time Analysis for Fetching Various Features.}
\begin{tabular} {|c|c|}
\hline
{\bf Feature type} & {\bf Average fetch time}\\  
\hline
HTML + JavaScript & 3.1 s / link\\
ECDF & 1.9 s / link\\
Topic + translation & 3.4 s / link\\
Topic + without translation & 1.3 s / link\\
\hline
\end{tabular}
\label{tbl:timePerformance}
\vspace{-3mm}
\end{table}

\subsection{Limitations}
\iffalse
We have identified three general limitations of our work: the subjectivity of crowdsourced webpage reputation systems, the completeness of available features, and the potential vulnerability to adversarial manipulation of LookAhead.
Firstly, there are potential issues with using any kind of crowdsourced reputation system ranging from subjectivity of ratings to manipulation attacks against such systems. In the case of WOT another limitation is that parts of the rating system are proprietary. In the case of LookAhead these mainly introduce uncertainty into the prediction and make the classification task more challenging. However, the separation into bad and good websites helps reduce the impact of subjective personal opinion.
Secondly, LookAhead achieves the best results when all possible features are available, but this is not always the case. In case of partial features the performance degrades. This could partially be suppressed by expanding the feature-space to consider additional content-based features in addition to outgoing links and topics.
\fi
Perhaps the most significant limitation of any system using machine learning to detect bad websites is the potential for adversaries to manipulate the system: either by modifying their website to avoid detection or by manipulating the classifier itself.
While the use of the ECDF-function protects against manipulation of outgoing links, as we pointed out in
Section~\ref{sssec:topic}, the simplistic approach of using topic
modeling is vulnerable to an attacker who attempts to influence the
inferred topic model for a page he controls. Instead of directly using
the probability distribution of topics as we do in
Section~\ref{sssec:topic}, we could convert to a boolean vector
(indicating if the topic is present on the page). Such an approach
will reduce FN (since an attacker can no longer gain by adding text to
his page to make it appear to belong to an innocuous topic as the
dominant topic), but will also raise FPs. We are currently
investigating this avenue.

Another limitation is that, although the performance of LookAhead is comparable to previous solutions, real time use will require further speedup. One option here is to use server-side assisted feature extraction.
Finally,
an open question is how the use of predicted ratings will
influence the actual rating. For example, if the predicted rating is
used for intermediate user feedback as suggested above, it might sway
future input ratings from the crowd towards the predicted rating.

\subsection{Current Work} 
We are conducting a
longitudinal study on a large number of websites that do not yet have
a WOT rating. We plan to see (a) how well our predictions match those
websites that do eventually get a rating and (b) how do our
predictions as well as the actual ratings evolve over time.

%\balance

% - Analysis: results
% - Analysis: features/webpages
% - Performance

%%%%%%%% Acknowledgements %%%%%%%
\section{Acknowledgements}
This work was partially supported
by the Intel Institute for Collaborative Research in Secure Computing
(ICRI-SC) and the Academy of Finland project ``Contextual Security'' (Grant Number: 274951).
We thank Web of Trust for giving access to their data which we used in
this work, Timo Ala-Kleemola and Sergey Andryukhin for helping us understand the WOT data,
Jian Liu and Swapnil Udar for helping to develop the web crawler.
We would also like to thank Petteri Nurmi, Pekka Parviainen, and Nidhi Gupta for their
feedback on an earlier version of this manuscript.

\bibliographystyle{abbrv}

\begin{thebibliography}{10}

\bibitem{AF13}
D.~Akhawe and A.~P. Felt.
\newblock Alice in warningland: A large-scale field study of browser security
  warning effectiveness.
\newblock In {\em Proceedings of the 22Nd USENIX Conference on Security},
  SEC'13, pages 257--272, Berkeley, CA, USA, 2013. USENIX Association.

\bibitem{anderson2007spamscatter}
D.~S. Anderson, C.~Fleizach, S.~Savage, and G.~M. Voelker.
\newblock {\em Spamscatter: Characterizing internet scam hosting
  infrastructure}.
\newblock PhD thesis, University of California, San Diego, 2007.

\bibitem{bhattacharya14using}
S.~Bhattacharya, P.~Nurmi, N.~Hammerla, and T.~Pl{\"o}tz.
\newblock Using unlabeled data in a sparse-coding framework for human activity
  recognition.
\newblock {\em Pervasive and Mobile Computing}, May 2014.

\bibitem{bishop07pattern}
C.~M. Bishop.
\newblock {\em Pattern Recognition and Machine Learning}.
\newblock Springer, 2007.

\bibitem{blei12probabilistic}
D.~M. Blei.
\newblock Probabilistic topic models.
\newblock {\em Communications of the ACM}, 55(4):77--84, 2012.

\bibitem{blei03latent}
D.~M. Blei, A.~Y. Ng, and M.~I. Jordan.
\newblock Latent dirichlet allocation.
\newblock {\em Journal of Machine Learning Research}, 3:993--1022, 2003.

\bibitem{breese98empirical}
J.~S. Breese, D.~Heckerman, and C.~Kadie.
\newblock Empirical analysis of predictive algorithms for collaborative
  filtering.
\newblock In {\em Proceedings of the Fourteenth Conference on Uncertainty in
  Artificial Intelligence}, pages 43--52, 1998.

\bibitem{breiman2001random}
L.~Breiman.
\newblock Random forests.
\newblock {\em Machine learning}, 45(1):5--32, 2001.

\bibitem{canali11prophiler}
D.~Canali, M.~Cova, G.~Vigna, and C.~Kruegel.
\newblock Prophiler: A fast filter for the large-scale detection of malicious
  web pages.
\newblock In {\em Proceedings of the 20th International Conference on World
  Wide Web}, pages 197--206. ACM, 2011.

\bibitem{CK11}
P.~H. Chia and S.~J. Knapskog.
\newblock Re-evaluating the wisdom of crowds in assessing web security.
\newblock In G.~Danezis, editor, {\em Financial Cryptography}, volume 7035 of
  {\em Lecture Notes in Computer Science}, pages 299--314. Springer, 2011.

\bibitem{cova2010detection}
M.~Cova, C.~Kruegel, and G.~Vigna.
\newblock Detection and analysis of drive-by-download attacks and malicious
  javascript code.
\newblock In {\em Proceedings of the 19th international conference on World
  wide web}, pages 281--290. ACM, 2010.

\bibitem{cox84analysis}
D.~R. Cox and D.~Oakes.
\newblock {\em Analysis of Survival Data}.
\newblock Champman and Hall, CRC, 1984.

\bibitem{curtsinger2011zozzle}
C.~Curtsinger, B.~Livshits, B.~G. Zorn, and C.~Seifert.
\newblock Zozzle: Fast and precise in-browser javascript malware detection.
\newblock In {\em USENIX Security Symposium}, pages 33--48, 2011.

\bibitem{daigle2004whois}
L.~Daigle.
\newblock Whois protocol specification.
\newblock \\http://tools.ietf.org/html/rfc3912, 2004.

\bibitem{whois}
L.~Daigle.
\newblock Rfc 3912: Whois protocol specification, September 2014.
\newblock http://tools.ietf.org/html/rfc3912.

\bibitem{FP07}
B.~Feinstein and D.~Peck.
\newblock {Caffeine Monkey: Automated Collection, Detection and Analysis of
  Malicious JavaScript}.
\newblock In {\em In Proceedings of the Black Hat Security Conference, 2007},
  2007.

\bibitem{Felegyhazi10}
M.~Felegyhazi, C.~Kreibich, and V.~Paxson.
\newblock On the potential of proactive domain blacklisting.
\newblock In {\em Proceedings of the 3rd USENIX Conference on Large-scale
  Exploits and Emergent Threats: Botnets, Spyware, Worms, and More}, LEET'10,
  pages 6--6, Berkeley, CA, USA, 2010. USENIX Association.

\bibitem{fukunaga90introduction}
K.~Fukunaga.
\newblock {\em Introduction to Statistical Pattern Recognition}.
\newblock Academic Press, second edition, 1990.

\bibitem{hammerla13preserving}
N.~Hammerla, R.~Kirkham, P.~Andras, and T.~Pl\"otz.
\newblock On preserving statistical characteristics of accelerometry data using
  their empirical cumulative distribution.
\newblock In {\em Proceeding of International Symposium on Wearable Computers
  (ISWC)}, 2013.

\bibitem{kittler98combining}
J.~Kittler, M.~Hatef, R.~Duin, and J.~Matas.
\newblock On combining classifiers.
\newblock {\em IEEE Transactions on Pattern Analysis and Machine Intelligence},
  20(3):226--239, 1998.

\bibitem{LJO09}
P.~Likarish, E.~Jung, and I.~Jo.
\newblock Obfuscated malicious javascript detection using classification
  techniques.
\newblock In {\em 4th International Conference on Malicious and Unwanted
  Software (MALWARE)}, pages 47--54, 2009.

\bibitem{liu05toward}
H.~Liu and L.~Yu.
\newblock Toward integrating feature selection algorithms for classification
  and clustering.
\newblock {\em IEEE Transactions on Knowledge and Data Engineering},
  17(4):491--502, 2005.

\bibitem{Ma09BeyondBlacklists}
J.~Ma, L.~K. Saul, S.~Savage, and G.~M. Voelker.
\newblock Beyond blacklists: Learning to detect malicious web sites from
  suspicious {URL}s.
\newblock In {\em Proceedings of the 15th ACM SIGKDD International Conference
  on Knowledge Discovery and Data Mining}, KDD '09, pages 1245--1254, New York,
  NY, USA, 2009. ACM.

\bibitem{mcnemar47note}
Q.~McNemar.
\newblock Note on the sampling error of the difference between correlated
  proportions or percentages.
\newblock {\em Psychometrika}, 12:153--157, 1947.

\bibitem{menardi14training}
G.~Menardi and N.~Torelli.
\newblock Training and assessing classification rules with imbalanced data.
\newblock {\em Data Mining and Knowledge Discovery}, 28(1):92--122, 2014.

\bibitem{tyler08evaluating}
T.~Moore and R.~Clayton.
\newblock Evaluating the wisdom of crowds in assessing phishing websites.
\newblock In {\em Financial Cryptography and Data Security}. Springer Berlin
  Heidelberg, 2008.

\bibitem{MC09}
T.~Moore and R.~Clayton.
\newblock Temporal correlations between spam and phishing websites.
\newblock In {\em Proceedings of the 2nd USENIX Conference on Large-scale
  Exploits and Emergent Threats: Botnets, Spyware, Worms, and More (LEET)},
  pages 5--5, 2009.

\bibitem{plotz11feature}
T.~Pl\"otz, N.~Y. Hammerla, and P.~Olivier.
\newblock Feature learning for activity recognition in ubiquitous computing.
\newblock In {\em International Joint Conference on Artificial Intelligence
  (IJCAI)}, pages 1729--1734, 2011.

\bibitem{Prakash10}
P.~Prakash, M.~Kumar, R.~Kompella, and M.~Gupta.
\newblock Phishnet: Predictive blacklisting to detect phishing attacks.
\newblock In {\em INFOCOM, 2010 Proceedings IEEE}, pages 1--5, March 2010.

\bibitem{rieck2010cujo}
K.~Rieck, T.~Krueger, and A.~Dewald.
\newblock Cujo: efficient detection and prevention of drive-by-download
  attacks.
\newblock In {\em Proceedings of the 26th Annual Computer Security Applications
  Conference}, pages 31--39. ACM, 2010.

\bibitem{wotstat}
J.~Ruvolo.
\newblock {WOT} statistics, Dec 2014.
\newblock https://www.mywot.com/en/community/statistics.

\bibitem{sagha11benchmarking}
H.~Sagha, S.~Digumarti, J.~del R~Millan, R.~Chavarriaga, A.~Calatroni,
  D.~Roggen, and G.~Tr\"oster.
\newblock Benchmarking classification techniques using the {Opportunity} human
  activity dataset.
\newblock In {\em IEEE International Conference on Systems, Man, and
  Cybernetics (SMC)}, 2011.

\bibitem{seifert08identification}
C.~Seifert, I.~Welch, and P.~Komisarczuk.
\newblock Identification of malicious web pages with static heuristics.
\newblock In {\em Telecommunication Networks and Applications Conference
  (ATNAC)}, pages 91--96, 2008.

\bibitem{seifert08identify}
C.~Seifert, I.~Welch, P.~Komisarczuk, C.~Aval, and B.~Popovsky.
\newblock Identification of malicious web pages through analysis of underlying
  dns and web server relationships.
\newblock In {\em 33rd IEEE Conference on Local Computer Networks (LCN)}, pages
  935--941, 2008.

\bibitem{truong14company}
H.~T.~T. Truong, E.~Lagerspetz, P.~Nurmi, A.~J. Oliner, S.~Tarkoma, N.~Asokan,
  and S.~Bhattacharya.
\newblock The company you keep: Mobile malware infection rates and inexpensive
  risk indicators.
\newblock In {\em Proceedings of the 23rd International Conference on World
  Wide Web}, pages 39--50, 2014.

\bibitem{voyant}
{Voyant Tools}.
\newblock {\em {VOYANT - See through your text}}, Nov 2014.
\newblock http://voyant-tools.org/.

\end{thebibliography}

%\input{chapters/appendices}

%%%%%%%% Discussion %%%%%%%
\ifsubmission
\else
%!TEX root = submission.tex
\newpage
\appendix

\subsection{Topic Modeling}
\label{app:topic}

\begin{figure}[!h]
\centering
\includegraphics[width=0.65\linewidth]{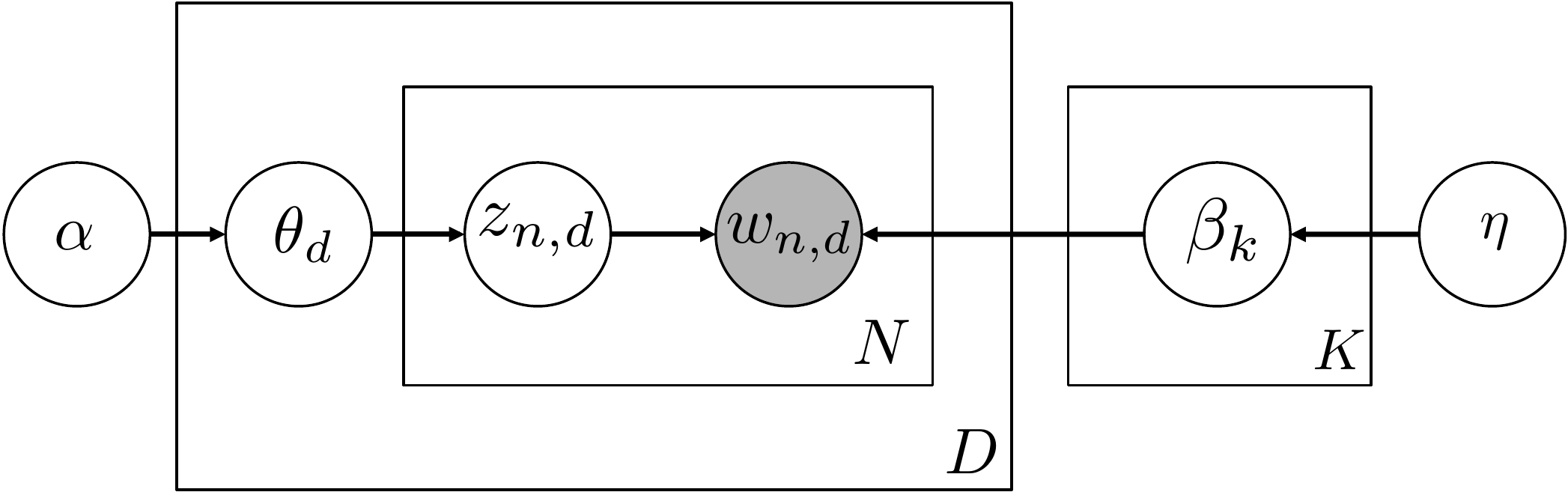}
\caption{Graphical model representing the LDA used to analyze text form web pages. The boxes represent replications, the shaded circle is the observed variable (word) and unshaded circles are unobserved variables.}
\label{fig:LDA}
\end{figure}

A graphical representation of the model that we use in our analysis of web texts is shown in Figure~\ref{fig:LDA}. 
%LDA follows a probabilistic generative modeling approach, where documents, in our case web pages containing text, are represented as a mixture of $k$ latent (unobserved) topics, where each topic is characterized by a distribution over a fixed set of words or vocabulary $V$. 
LDA follows a probabilistic generative modeling approach using the {\em bag-of-words} assumption, i.e., the order of occurrence of words within a document is ignored. Under the LDA model, each web page or document $d \in D$  is represented as a mixture of $k$ latent (unobserved) topics denoted by the multinomial variable $\theta_d$ (topic proportion of the $d^{th}$ document), where each topic $\beta_k$ is a distribution over the set of words or vocabularies $V$ and is sampled from a Dirichlet distribution with parameter $\eta$. $\theta_d$ is also drawn from a Dirichlet distribution with parameter $\alpha$. $z_{n,d}$ denotes the topic-assignment for the observed word $w_{n,d}$, which is sampled from $\theta_d$. Each word $w_{n,d}$ depends on the topic-assignment variable $z_{n,d}$ and all the topic distributions $\{\beta_k\}_{k=1}^K.$ The joint probability distribution can be written as~\cite{blei12probabilistic}:
\begin{eqnarray}
p(\beta_{1:K}, \theta_{1:D}, z_{1:D}, w_{1:D}) & =  \nonumber \\ 
\prod_{k=1}^K p(\beta_k | \eta) \prod_{d=1}^D p(\theta_d | \alpha) \nonumber \\
\left(  \prod_{n=1}^N p(z_{n,d} | \theta_d) p(w_{n,d} | z_{n,d}, \beta_{1:K})\right) &
\end{eqnarray}

The main task of the topic model is thus to infer the parameters $\beta_k, \theta_d$ and $z_{n,d}$ from the corpus of text, i.e., $w_{1:D}$. The posterior distribution can be written as:
\begin{eqnarray}
p(\beta_{1:K}, \theta_{1:D}, z_{1:D} |  w_{1:D}) = \frac{p(\beta_{1:K}, \theta_{1:D}, z_{1:D}, w_{1:D})}{p(w_{1:D})} \label{eqn:topic}
\end{eqnarray}
Although, there are variational algorithms proposed in the literature for estimating the posterior distribution given in Equation~\ref{eqn:topic}, we use a Gibbs sampling-based approach to efficiently approximate it. Once the model parameters are estimated, for a given web page or document $d$ containing the set of words\footnote{As a common pre-processing step, we remove all frequently occurring words or stop words (in the language) from a document.} $w$, we use $p(\theta_d | w, \beta_{1:K})$, i.e., the estimated topic proportion as the feature set for capturing the thematic content of the document.

With topic modeling we gain insights into the basic composition of the web content present in our opportunistic dataset. The most frequent topics encountered were related to commenting and sharing, adult content, financial topics, and gaming. The most frequently occurring words are visualized in Fig.~\ref{fig:topic_visual}, which is created using the online tool Voyant~\cite{voyant}. In the figure the size of individual words reflects its relative number of occurrences in our sample set of web pages.

\begin{figure}[!t]
\centerline{
	\label{fig:topics}
	\includegraphics[width=0.65\linewidth]{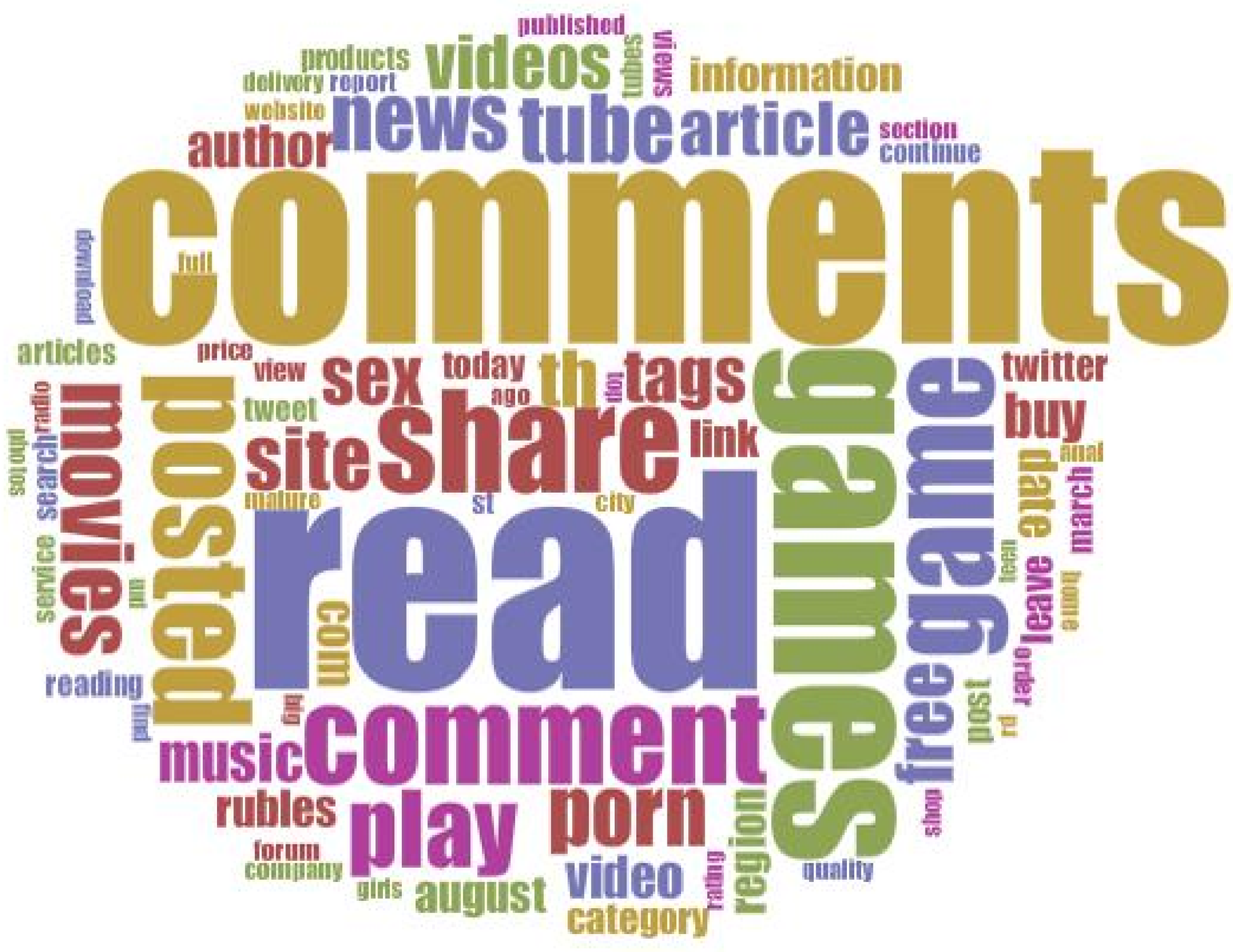}\label{fig:topics}}
\vspace{-2mm}	
\caption{Textual summary of the most frequent words present in our opportunistic dataset, where the size indicated the frequency of the words in the web text.}
\label{fig:topic_visual}
\vspace{-3mm}
\end{figure}

\subsection{Fukunaka Class Separability}
\label{app:fukunaka}

The Fukunaga score relies on computing both the {\em within-class} and {\em between-class} scatter matrices. Let $n_i$ be the number of samples of the feature set belonging to class $c_i$ of $C$ different classes. Further, let $\bm{\mu}_i$ and $\bm{\mu}$ respectively be the mean for class $c_i$ and the global mean. The within-class and between-class scatter matrices $S_W$ and $S_B$ can be computed as follows:
\begin{eqnarray}
S_W &=& \sum_{i=1}^C \left[ \sum_{j=1}^{n_i} (\bm{f}_{i,j} - \bm{\mu}_i) (\bm{f}_{i,j} - \bm{\mu}_i)^T\right]\\
S_B &=& \sum_{i=1}^C n_i (\bm{\mu}_i - \bm{\mu}) (\bm{\mu}_i - \bm{\mu})^T
\end{eqnarray}
Then the Fukunaga class separability score is computed as:
\begin{eqnarray} 
Separability = Tr(S_B / S_W)
\end{eqnarray}
To illustrate the effect of classification combination strategy, Fig.~\ref{fig:classifierWeights} shows the average (computed over 10 folds) relative weights, computed using Fukunaga class separability, as used by \sys{} for the all-valid dataset.

\begin{figure}[!t]
\centering
\includegraphics[width=0.65\linewidth]{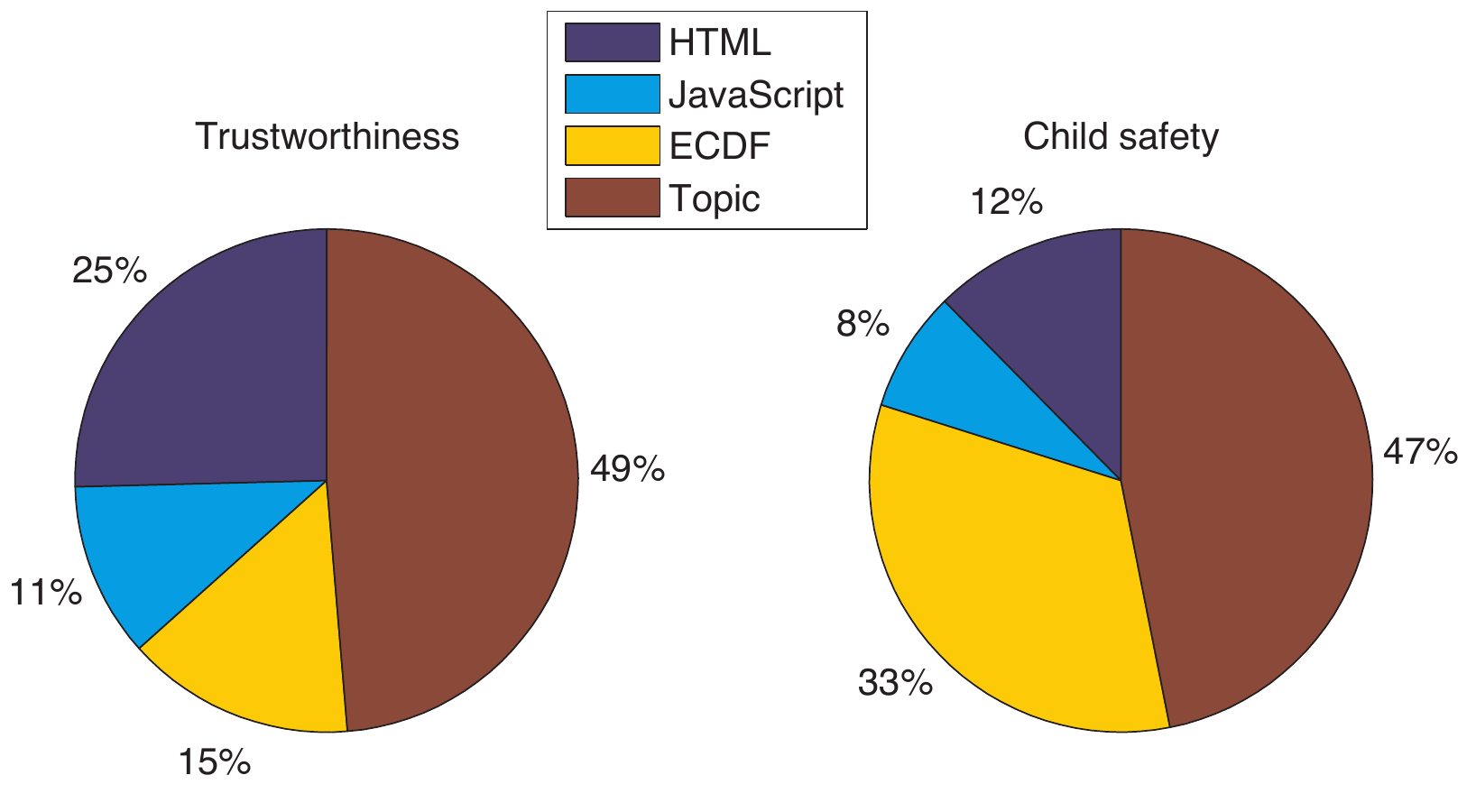}
\vspace{-3mm}
\caption{Average classifier combination weights used by \sys{} on the all-valid dataset.}
\label{fig:classifierWeights}
\vspace{-3mm}
\end{figure}

\subsection{Classifier Combination Rules}
\label{app:rules}

\subsubsection{Sum Rule}
\begin{eqnarray} 
p^*(c_j | \bm{f}_H, \bm{f}_J, \bm{f}_E, \bm{f}_T) = \sum_{k \in \{H,J,E,T\}} p(c_j | \bm{f}_k)
\label{eqn:map} 
\end{eqnarray}

\subsubsection{Product Rule}
\begin{eqnarray} 
p^*(c_j | \bm{f}_H, \bm{f}_J, \bm{f}_E, \bm{f}_T) = \prod_{k \in \{H,J,E,T\}} p(c_j | \bm{f}_k)
\label{eqn:map} 
\end{eqnarray}

\subsubsection{OR Rule}
If at least one of the constituent classifiers predicts a URL to be bad, then the overall prediction is bad. 

\subsubsection{Majority Voting}
The final classification is based on the majority voting of the constituent classifiers. Ties are broken randomly.

\begin{table}[!t]
\centering
\begin{threeparttable}
\caption{Performance of \sys, under various feature combinations, and Prophiler on the all-valid dataset.}
\begin{tabular} {|c|c|c|c|c|c|c|}
\multicolumn{7}{l}{\bf All-valid dataset size: 43,675 URLs,  $\bm{\mathcal{T}_h} = 60$, $C_r = 60$}\\
\multicolumn{7}{l}{\bf Reputation dimension: Trustworthiness}\\
\hline
\multicolumn{4}{|c|}{\bf Feature sets}      &  {\bf Avg. F$_1$-Score} & {\bf FNR} & {\bf FPR}\\  
\cline{1-4}
{\bf H} & {\bf J} & {\bf E} & {\bf T} & (\%) & (\%) & (\%)\\
\hline
\checkmark &  &  &   & $75.7$ **& $23.6$ & $24.9$\\
& \checkmark &  &  &  $74.9$ **& $25.1$ & $25.1$\\
& & \checkmark &  &  $70.5$ **& $29.0$ & $29.9$\\
&  & & \checkmark &  $74.9$ **& $24.5$ & $25.6$\\
\hline
\checkmark & \checkmark &  &  &  $76.8$ **& $22.6$ & $23.8$\\
\checkmark &  & \checkmark &  &  $77.2$ **& $24.0$ & $21.6$\\
\checkmark &  &  & \checkmark &  $78.7$ **& $20.1$ & $22.3$\\
& \checkmark & \checkmark &  &  $74.1$ **& $26.4$ & $25.5$\\
& \checkmark &  & \checkmark &  $77.8$ **& $21.7$ & $22.7$\\
&  & \checkmark & \checkmark &  $79.1$ **& $21.1$ & $20.6$\\
\hline
\checkmark & \checkmark &  \checkmark &  &  $79.2$ **& $21.7$ & $19.8$\\
\checkmark &  \checkmark &  & \checkmark &  $79.6$ **& $19.5$ & $21.2$\\
\checkmark &  & \checkmark & \checkmark &  $81.7$ **& $18.0$ & $18.7$\\
&  \checkmark & \checkmark & \checkmark &  $80.9$ **& $19.1$ & $19.1$\\
\hline
\rowcolor{Gray}
\checkmark & \checkmark & \checkmark & \checkmark &  \bm{$82.4$} & \bm{$17.3$} & \bm{$17.9$}\\
\hline
\multicolumn{7}{l}{}\\
\hline
\multicolumn{4}{|c|}{\bf Prophiler} &  $75.3$ **& $15.0$ & $33.7$\\
\hline

\multicolumn{7}{l}{}\\
\multicolumn{7}{l}{\bf All-valid dataset size: 42,334 URLs,  $\bm{\mathcal{T}_h} = 60$, $C_r = 60$}\\
\multicolumn{7}{l}{\bf Reputation dimension: Child safety}\\
\hline
\multicolumn{4}{|c|}{\bf Feature sets}      &  {\bf Avg. F$_1$-Score} & {\bf FNR} & {\bf FPR}\\  
\cline{1-4}
{\bf H} & {\bf J} & {\bf E} & {\bf T} & (\%) & (\%) & (\%)\\
\hline
 \checkmark &  &  &   & $79.2$ **& $16.7$ & $26.4$\\
& \checkmark &  &  &  $78.7$ **& $16.5$ & $27.9$\\
& & \checkmark &  &  $73.4$ **& $22.4$ & $32.4$\\
&  & & \checkmark &  $80.9$ **& $17.9$ & $20.8$\\
\hline
\checkmark & \checkmark &  &  &  $79.9$ **& $15.7$ & $26.0$\\
\checkmark &  & \checkmark &  &  $79.0$ **& $17.4$ & $25.8$\\
\checkmark &  &  & \checkmark &  $83.2$ **& $14.7$ & $19.7$\\
& \checkmark & \checkmark &  &  $76.8$ **& $19.4$ & $28.5$\\
& \checkmark &  & \checkmark &  $82.8$ **& $15.6$ & $19.6$\\
&  & \checkmark & \checkmark &  $83.1$ **& $15.8$ & $18.6$\\
\hline
\checkmark & \checkmark &  \checkmark &  &  $81.4$ **& $14.8$ & $23.6$\\
\checkmark &  \checkmark &  & \checkmark &  $84.0$ **& $13.4$ & $19.6$\\
\checkmark &  & \checkmark & \checkmark &  $84.8$ **& $13.5$ & $17.5$\\
&  \checkmark & \checkmark & \checkmark &  $84.3$ **& $14.3$ & $17.8$\\
\hline
\rowcolor{Gray}
\checkmark & \checkmark & \checkmark & \checkmark &  \bm{$85.3$} & \bm{$12.8$} & \bm{$17.4$}\\
\hline
\multicolumn{7}{l}{}\\
\hline
\multicolumn{4}{|c|}{\bf Prophiler} &  $78.4$ **& $10.5$ & $35.8$\\
\hline
\end{tabular}
\label{tbl:allvalidth60}
\begin{tablenotes}
\footnotesize
\item **: Statistically significant with $99\%$ confidence.
%\item * : Statistically significant with $95\%$ confidence.
\end{tablenotes}
\end{threeparttable}
\vspace{-1mm}
\end{table}

\subsection{Evaluation Metrics}
\label{app:fpr}

In our experimental evaluations we use F$_1$-score (expressed as a percentage) as the main performance indicator, which is computed as: 
\begin{eqnarray}
\mbox{F$_1$-score} &=& \frac{2 \cdot \mbox{precision} \cdot \mbox{recall}}{\mbox{precision} + \mbox{recall}},\,\,\mbox{where}\\
\mbox{precision} &=& \frac{TP}{TP + FP},\\
\mbox{and}\,\, \mbox{recall} &=& \frac{TP}{TP + FN}.
\end{eqnarray}
Here, TP = True positives, FP = False positives, and FN = False negatives.

In line with standard practices~\cite{bhattacharya14using,sagha11benchmarking}, to overcome non-uniform class distribution in the test dataset, we use the weighted average of the individual F$_1$-score of all classes.   
\begin{eqnarray}
\mbox{Avg. F$_1$-score} =  \frac{\sum_{i=1}^c w_i \cdot F_1^i\mbox{-score}}{\sum_{i=1}^c w_i},
\end{eqnarray}
where, $F_1^i\mbox{-score}$ is the F$_1$-score of the $i^{th}$ class and $w_i$ is the number of samples of class $i$ in the test dataset. Additionally, we report the FNR and FPR, computed from the confusion matrix as:

\begin{eqnarray}
FNR = \frac{FN}{FN + TP}\,,\,\,\,FPR = \frac{FP}{FP + TP},
\end{eqnarray}
%where,  FN = False Negatives, TP = True Positives and FP = False Positives.

\subsection{Results}

Similarly to the Section~\ref{sec:eval}, for completeness, we present performance of \sys{} and Prophiler on the all-valid dataset for $\bm{\mathcal{T}_h} = 60$, $C_r = 60$ in Table~\ref{tbl:allvalidth60}.
Note that, a higher value of $C_r$ allows more URLs to have valid ECDF features and thus increases the dataset sizes in both dimensions compared to the results presented in Table~\ref{tbl:allvalidth40}. Table~\ref{tbl:opportunisticth60} presents the performance of \sys{} and Prophiler on the opportunistic dataset with $\bm{\mathcal{T}_h} = 60$, $C_r = 60$, while using various classifier combination rules.

\begin{table}[!t]
\centering
\begin{threeparttable}
\caption{Performance of \sys{} on the opportunistic dataset.}
\begin{tabular} {|c|c|c|c|c|}
\multicolumn{5}{l}{\bf Opportunistic dataset size: 89,220 URLs, $\bm{\mathcal{T}_h}=60$ and $C_r=60$}\\
\multicolumn{5}{l}{\bf Reputation dimension: Trustworthiness}\\
\hline
\multicolumn{2}{|c|}{\bf Experiment} & {\bf Avg. F$_1$-Score} & {\bf FNR} & {\bf FPR}\\  
\cline{1-4}
{\bf Comb. Rule} & {\bf Balancing} & (\%) & (\%) & (\%)\\
\hline
\rowcolor{Gray}
Adaptive & & \bm{$80.3$} & $33.9$ & $10.5$\\
Sum & & $79.6$ \mbox{ *} & $38.0$ & \bm{$8.7$}\\
Product & & $80.0$ \mbox{ *}& $36.0$ & $9.6$\\
Or & & $44.3$ ** & \bm{$5.0$} & $77.6$\\
Voting & & $71.4$ **& $23.6$ & $32.4$\\
\hline
Prophiler* & & $71.7$ ** & $30.5$ & $27.5$\\
\hline
\hline
\rowcolor{Gray}
Adaptive & \checkmark & \bm{$77.1$} & $20.0$ & \bm{$25.2$}\\
Sum & \checkmark & $77.0$ & $19.4$ & $25.7$\\
Product & \checkmark & $76.6$ **& $20.0$ & $26.1$\\
Or & \checkmark & $40.5$ **& \bm{$1.9$} & $82.4$\\
Voting & \checkmark & $64.2$ **& $10.5$ & $50.3$\\
\hline
Prophiler* & \checkmark & $65.8$ ** & $14.6$ & $46.2$\\
\hline
\multicolumn{5}{l}{}\\
\multicolumn{5}{l}{\bf Opportunistic dataset size: 84,714 URLs, $\bm{\mathcal{T}_h}=60$ and $C_r=60$}\\
\multicolumn{5}{l}{\bf Reputation dimension: Child safety}\\
\hline
\multicolumn{2}{|c|}{\bf Experiment}      &  {\bf Avg. F$_1$-Score} & {\bf FNR} & {\bf FPR}\\  
\cline{1-4}
{\bf Comb. Rule} & {\bf Balancing} & (\%) & (\%) & (\%)\\
\hline
\rowcolor{Gray}
Adaptive & & \bm{$82.6$} & $26.8$ & $10.0$\\
Sum & & $82.4$ \mbox{ *} & $28.7$ & \bm{$8.9$}\\
Product & & $82.3$ \mbox{ *} & $27.4$ & $10.1$\\
Or & & $45.4$ **& \bm{$4.4$} & $80.0$\\
Voting & & $73.0$ ** & $17.7$ & $34.0$\\
\hline
Prophiler* & & $72.5$ ** & $23.9$ & $30.4$\\
\hline
\hline
\rowcolor{Gray}
Adaptive & \checkmark & \bm{$80.8$} & $25.9$ & \bm{$14.1$}\\
Sum & \checkmark & $80.6$ \mbox{ *}& $23.6$ & $16.3$\\
Product & \checkmark & $80.3$ **& $23.7$ & $16.7$\\
Or & \checkmark & $45.3$ **& \bm{$3.6$} & $80.3$\\
Voting & \checkmark & $69.3$ **& $13.0$ & $43.3$\\
\hline
Prophiler* &\checkmark & $69.7$ **& $16.0$ & $40.7$\\
\hline
\end{tabular}
\label{tbl:opportunisticth60}
\begin{tablenotes}
\footnotesize
\item **: Statistically significant with $99\%$ confidence.
\item * : Statistically significant with $95\%$ confidence.
\end{tablenotes}
\end{threeparttable}
\end{table}

\fi

% that's all folks
\end{document}